\documentstyle[12pt,aasms4]{article}

\received{1995 December 15}
\accepted{26 April 1996}

\slugcomment{To appear in The Astronomical Journal, August 1996 issue}

\lefthead{M\'endez}
\righthead{Galactic structure models}

\begin{document}

\title{Galactic structure towards the open clusters NGC 188 and NGC 3680}

\author{R. A. M\'endez}
\affil{Yale University Observatory and European Southern
Observatory}
\author{W. F. van Altena}
\affil{Yale University Observatory}

\begin{abstract}

We present the first comparisons of a newly developed Galactic
Structure and Kinematic Model to magnitude and color
counts, as well as relative proper motions, in the fields of the open
clusters NGC 188 ($(l, b)=(122.8^o, +22.4^o)$) and NGC 3680
($(l,b)=(286.8^o, +16.9^o)$). In addition to determining the reddening
toward these two clusters, it is
shown that starcounts at intermediate Galactic latitudes in the
range $11 \le V \le 17$ allow us to constrain the model scale-height for disk
subgiants. We obtain a mean value of $250 \pm 32$ pc, in agreement with
previous determinations of the scale-height for {\it red-giants}. We
are also able to constrain the
scale-height of main-sequence stars, and the distance of the sun from
the Galactic plane, ruling out the possibility of a value of +40 pc, 
in favor of a smaller value. Comparisons with the observed
proper-motion histograms
indicate that the velocity dispersion of disk main-sequence stars must
increase with distance from the Galactic plane in order to match the
observed proper-motion dispersion. The required increase is consistent
with the values predicted by dynamical models, and provides a clear 
observational evidence in favor of such gradients. The {\it shape} of the
observed proper-motion distribution is well fitted within the Poisson
uncertainties. This implies that
corrections to absolute proper motion (and, therefore, space
velocities) for open clusters may be obtained using our
model when no inertial reference frame is available. Using this
approach, the derived tangential motions for NGC 188 and NGC 3680 are 
presented.

\end{abstract}


\section{Introduction}

The use of starcount models to constrain {\it global} Galactic Structure
parameters has proved to be an effective way of investigating the
broad properties of stellar populations in our Galaxy
(\cite{rema93} and references therein). However, most of the proposed
models have lacked the
ability to predict starcounts {\it and} kinematics simultaneously, with a
few notable exceptions (\cite{roob87,rbc89}). Furthermore, models
that do predict kinematics have not been extensively used. With the
wealth of information related to Galactic Structure and Kinematics that is
expected to emerge from the Northern Proper Motion Program
(\cite{klet87}), and its Southern counterpart, the Southern Proper Motion
Program (\cite{vaet94}), as well as from the ongoing
measurement of POSS-II plates (\cite{rh93}), we thought it would be 
interesting to explore the potentials and limitations of a galactic
model that incorporates kinematics, in preparation for the advent of
these mammoth photometric and kinematic surveys. An additional
motivation is the need for a code to predict secular proper motion
(\cite{ma93}) and the correction to absolute parallax (\cite{mon88})
for any position in the sky, and under arbitrary selection constraints. In
what follows we present the first comparisons of our model to
magnitude and color counts, as well as proper-motion distributions
for two surveys in the fields of the Galactic clusters NGC 188 and NGC
3680. In the second section a description of the model is provided.
Then, in the third section, a brief description of the data is
outlined and the model comparisons to star- and color counts as well as kinematic histograms is presented. Finally, in the fourth
section, a brief summary of our findings is given.

\section{The model}

\subsection{The starcounts model}

Starcounts are computed by using the so-called
{\it fundamental equation of stellar statistics} (\cite{tw62,mb81}).
We have followed a parametrization of the density functions,
luminosity functions, and Hess-diagrams, similar to that employed by other
recent starcount models (see, e.g., \cite{rema93}). We have considered a
three-component model that includes a disk, a thick-disk, and a halo. For our
present purposes of comparing to bright, low-latitude counts, the
contribution of thick-disk and halo stars is minimal, and can be
ignored. We have nevertheless included them for completeness and for future
applications of our model. It should be emphasized that {\it in no case} are
the present comparisons attempts to constrain any of the adopted
parameters for the thick-disk and halo, rather we concentrate on the
disk, which is the dominant component at least to $V \sim 20$ at the
Galactic Poles ({\cite{rema93}) and even at fainter magnitudes at low
Galactic latitudes. Table 1 summarizes the main parameters adopted in
the model. These parameters represent a compromise value between many
determinations, and should be viewed as free parameters to be constrained by
our comparisons to different data sets. Our task is to identify which
parameters have the most impact on the observables (Section 3.1 and 3.2).

Since the disk is the population targeted in this study, a brief description
of the adopted density function, luminosity function, and Hess-diagram is
necessary.
For our density function we have adopted a double-exponentially decaying layer away from the Galactic center in the
radial direction, and away from the Galactic plane in the direction
perpendicular to the Galactic plane. Surface photometry of edge-on
Spiral galaxies indicates that their radial ligth profile follows quite
closely that of an exponential function decaying away from the nucleus
(van der Kruit and Searle 1981). On the other hand, an isothermal,
self-gravitating disk exhibits a $sech^2(Z/H_z^s)$ density law
perpendicular to the Galactic plane (\cite{s42}) that approximates an 
exponential
function away from the plane (\cite{vks81}), where $H_z^s$ is the
corresponding $sech^2$ scale-height (see Equation 6). This has motivated the 
adoption of
various exponential-like functions as a {\it representation} of the stellar
density profile perpendicular to the Galactic plane for both the disk
and thick-disk (\cite{rema93,hb93,roet96,ojet96}), as well as
for individual tracers within these components (e.g., for planetary
nebulae \cite{zp91,md92}).

The stellar density of the Galactic
disk in different absolute magnitude intervals has been obtained by 
several authors. The latest value, determined
from the third edition of the Catalogue of Nearby Stars (CNS3
hereafter, \cite{jg93}), indicates a density of 0.12 $stars/pc^3$,
with a formal uncertainty of 10 \% (\cite{ja93}). This number has been
determined mainly from the stellar sample closer than 5 pc (66 stars),
where small number statistics can dominate over any true fluctuations 
in the stellar density. The sample might not be complete, for
example, if we also add reliable spectroscopic and astrometric
binaries, the number of stars within that distance limit increases to
70, which is already a change of 6 \% in the stellar density
(\cite{jg93}). Other
determinations of the stellar density are those of Agekjan and
Ogorodnikov (1974) which yield a value of $0.138 \pm 0.009 \:
stars/pc^3$ for $-1.0 \le M_v \le 19.5$ 
from the Catalogue of Nearby Stars by Woolley et
al. (1970), while from Wielen et al. (1983), we infer $0.119
\pm 0.012 \: stars/pc^3$ in the magnitude range $-1.5 \le M_v \le
21.5$. 
Indeed, one might expect this number to have fluctuations as lines of sight
cross regions of slightly different stellar density. However, Bahcall
(1986) has claimed this value to be constant within 15 \% in the range
$+5 \le M_v < +13.5$ from a comparison of his model to several
(high-latitude) fields. Bahcall's result is, therefore, in agreement with the
uncertainty of 10\% quoted by Jahreiss and Gliese from the CNS3, and
leaves very little room for big stellar density fluctuations in the
previously mentioned absolute magnitude range.


For the scale-length of the disk, we have adopted 3.5 kpc, similar to
other models of this kind (\cite{gi84,coet91,ba86}). It should be
noted however that some recent starcounts at low and intermediate
Galactic latitudes have lead to discrepant values; While
Robin et al. (1992) and Ojha et al. (1994a,b) found smaller values, in the range 
$\sim 2.0-2.5$ kpc (similar to the values inferred from infrared
starcounts, see \cite{keet91} for a review), Yamagata and Yoshii (1992) have found a value closer to 4.0 kpc
from starcounts in similar directions, although their counts do not seem to
agree with those by Soubiran (1992). We have adopted a variable scale-height for
main-sequence stars to account for the known fact that older stars have
diffused to larger distances from the Galactic plane than younger
stars (\cite{wifu83}). The functional form
described by Miller and Scalo (1979), and Bahcall and Soneira (1980),
which seems to be a good {\it representation} of the available observational
data (see also Gilmore and Reid 1983), has
been included in our model in the way indicated by Bahcall (1986). On
the other hand, subgiants, giants, and white-dwarfs have been assigned
a fixed value for the scale-height (see Table 1). Section 3.2 presents
the sensitivity of our model to both the scale-length and scale-height
of the disk. We find that
starcounts toward our two clusters are not sensitive to the
scale-length of the disk. The situation is different with the
scale-height: we are able to put significant constraints on this
parameter for subgiants and main-sequence stars (see Section 3.2).

Our disk luminosity function is Wielen's et al. (1983) for $M_v > +4$.
For  $M_v < -1$ Wielen's function becomes increasingly incomplete
(with only one star at $M_v = -1$, and 4 stars at $M_v = 0$ within 20
pc of the Sun), and
we have adopted instead McCuskey's function (1966) as a better
representation in this absolute magnitude range. There is very good 
agreement in the region of overlap, $-1 \le M_v < +4$, where we have
adopted the average
of both functions. This procedure is similar to that adopted by Bahcall et al.
(1987). Our present comparisons are not sensitive to bright giants and 
supergiants, but we have included this improved luminosity
function for future applications of the model. In addition, the still
large uncertainties at the faint end of the luminosity function ($M_v
> +10$, Kirkpatrick et al. 1994, Reid et al. 1995), do not affect the 
present results, which are mainly concerned
with stars in the range  $+1< M_v < +5$. Since the disk's luminosity
function has been well sampled in this magnitude range, we have
assumed it to be fixed and equal to the composite value derived in the
way described above.

As for the disk's Hess-diagram, we have adopted Robin and Cr\'ez\'e's
(1986) values, with the exception of the range $+2.75 \le M_v < 0.25$ and
$(B-V)_o \ge +0.9$ where
their diagram excludes some G-type subgiants. In this magnitude range
we have used the values derived from the CNS3 (there are some 40 stars
in the CNS3 that satisfy the above luminosity and color criteria). The lack
of subgiants in Robin
and Cr\'ez\'e's Hess-diagram is also indicated by a depression of about
a factor of two in their
luminosity function when compared to Wielen's et al. (1983) or
McCuskey's (1966) in the absolute magnitude range mentioned before. 
The origin of the discrepancy between the observed
density of subgiants in the local neighborhood and that adopted by Robin and Cr\'ez\'e is unclear,
but it has been recognized, indirectly, by Robin et al. (1992)
when comparing Robin and Cr\'ez\'e's model to low-latitude counts. A
detailed description of our method employed to update this region of the
Hess-diagram will be presented elsewhere as it bears minimal implications for the current application of our model.

\subsection{The kinematic model}\label{kine}

Velocity distributions along the line-of-sight and in the tangential
direction are computed {\it self-consistently} by using velocity ellipsoids
evaluated at each distance shell in the corresponding starcount
integration. We have adopted Schwarzschild's velocity ellipsoids
(1907, 1908). This velocity distribution {\it has not} been derived
dynamically, but it provides a good description of the observed
velocity distribution for stars in the solar neighborhood
({\cite{mb81,rbc89}). King (1990) has argued that it is reasonable to
assume that the velocity distribution takes a similar form elsewhere in
the Galaxy. However, it can be shown that Schwarzschild's hypothesis
predicts that the velocity dispersion in the radial direction is
independent of galactocentric distance (Fricke 1952), contrary to the
observations ({\cite{lf89}). On the other hand, equilibrium solutions
for low-velocity populations in disk-like orbits can indeed be well
approximated by Gaussians ({\cite{s69,vb85}). Also, as pointed out by
King (1990), Schwarzschild's velocity ellipsoids, despite their flaws,
lead to solutions that are able to predict many details of Galactic
structure and kinematics. Therefore, we have adopted the view that
these velocity ellipsoids are a good starting point for the kinematic
modeling. In what follows we describe the two main parameters that
specify the velocity ellipsoid, namely, velocity dispersions and velocity lags.

First-guess velocity dispersions have been taken from the literature.
Ratnatunga et al. (1989) have compiled results obtained by Delhaye
(1965). These values coincide approximately with the compilation
presented by Mihalas and Binney (1981), also derived from Delhaye's values. In addition, Ratnatunga et al. (1989) have derived a
set of velocity dispersions from a maximum likelihood analysis of the
Bright Star Catalogue (BSC hereafter, {\cite{ho82}), using the observed radial
velocities, {\it exclusively}. We have adopted a variation of the 
velocity dispersions with spectral type and
luminosity class similar to that of Ratnatunga et al. Our adopted velocity
dispersions are shown in Table 2. The velocity dispersions listed in
Table 2 are of course the {\it local} values appropriate for the solar
neighborhood. Unfortunately, there is very limited information
available on the {\it change} of velocity dispersion as a function of
position in the Galaxy (this is, of course, one of the motivations
for the present model). The only evidence comes from the study of K
giants by Lewis and Freeman (1989). They found that the component of
velocity dispersion along the radial direction (in a cylindrical
coordinate system) follows a decaying exponential function away from
the Galactic center, with a scale-length similar to that of the disk.
They also found that the V-component of the velocity dispersion (along
Galactic rotation) followed a similar trend, but with a different
scale-height, indicative of a non-flat rotation curve. In our model,
we have adopted this functional form for the U-velocity dispersion, while the
V-velocity dispersion has been assumed to follow from the
collisionless Boltzmann equation (CBE hereafter, {\cite{bt87}). As for
the vertical
velocity dispersion, $\Sigma_w$, dynamical balance requirements as well as
observations of disks in external galaxies ({\cite{bt87}) suggest
that $\Sigma_w$ should be proportional to the stellar (surface) density.
All these assumptions lead to the following functional variation for
our {\it model} velocity dispersions:

\begin{eqnarray}
\Sigma_u^2(R) & =  & \Sigma_u^2 \, e^{-\frac{(R-R_o)}{H_R}} \\
\Sigma_v^2(R) & =  & \frac{1}{2} \left( 1 + \frac{dlnV_c(R)}{dlnR} \right) \, \Sigma_u^2(R) \\
\Sigma_w^2(R) & =  & \Sigma_w^2 \, e^{-\frac{(R-R_o)}{H_R}}
\end{eqnarray}

where $R_o$ is the Solar galactocentric distance, $H_R$ is the disk
scale-length, $\Sigma_u$ and $\Sigma_w$ are the local velocity
dispersion in U and W respectively, and $V_c(R)$ is the circular speed
(i.e., the rotation curve or, equivalently, the motion of the Local
Standard of Rest) at distance R from the Galactic center (assumed to
have a local value of 220 km/s). The
Z-dependence on $\Sigma_w^2(R)$ has been removed by the vertical
integration of our adopted density law. Indeed, it seems that the vertical
gradient of the W-velocity dispersion is rather small
(\cite{fw87}), or zero (\cite{sl91}). We have started by assuming this
gradient to be zero for all three components of the velocity
dispersion (\cite{ad91}), but see Section 3.3.


The velocity ellipsoids are computed with respect to the mean speed of the
particular Galactic component being evaluated. Since, for an equilibrium
system, less rotational
support is needed for higher velocity dispersion stars (the so-called
asymmetric drift), we have evaluated the
mean speed from a self-consistent solution to the CBE (\cite{bt87}).
For a steady-state disk that is not expanding nor contracting, the
mean rotational speed, $<V(R,Z)>$, is given by

\begin{equation}
<V(R,Z)> = \sqrt{ V_c^2(R) - \Sigma_v^2(R)+ \left( 1 - \frac{2R}{H_R}
+ S(R,Z) \right) \, \Sigma_u^2(R) }
\end{equation}

where S(R,Z) is a function that describes the contribution to the
rotational support from the cross term $\Sigma_{uv}$ (usually referred
to as the tilt of the velocity ellipsoid). It can be shown that the
function S(R,Z) is given by

\begin{equation}
S(R,Z) = q \frac{(\lambda^2-1)R^2}{(R^2 + \lambda^2 Z^2)\lambda^2}
\left( \frac{(R^2 - \lambda^2 Z^2)}{(R^2 + \lambda^2 Z^2)} -
\frac{|Z|}{H_Z}) \right)
\end{equation}

where q is zero if the velocity ellipsoid has cylindrical symmetry, or one
if the velocity ellipsoid has spherical symmetry, $\lambda$ is the (fixed)
aspect ratio of the velocity ellipsoid,
defined by $\Sigma_u/\Sigma_w$, evaluated at Z= 0, and $H_z$ is the 
exponential
scale-height for the population considered. Equation (5)
has been obtained by using the expression for $\Sigma_{uv}$ presented by
Kuijken and Gilmore (1989a). The value for $\lambda$ is in the range 
1.5 to 2 for disk stars, e.g., Gilmore et al. (1989) adopted
$\lambda= 1.95$ for K and M dwarfs (see also \cite{ad91}), in close
agreement with our adopted value of $\lambda= 2.0$ for dwarfs fainter
than $M_v = +4.8$ (Table 2). The velocity lags computed from Equation
(4) are consistent with the values derived by Ratnatunga et al. (1989)
from their analysis of the BSC, as well as with those adopted by Robin
and Oblak (1987) in their kinematic model.

\section{The Data and Comparison to the Model}

\subsection{Overview}

We have selected two recent photometric and proper-motion surveys to
carry out our model comparisons. These are surveys in the fields of the
open clusters NGC 188 (\cite{di95}) and NGC 3680 (\cite{kp95}) whose
main objective was to provide membership probabilities. The
scope and accuracy of both surveys is quite similar, and are described
in the respective papers. The existence of clusters in the field of
view makes an analysis of the field stars more complex. However, we
gain from the astrometric point of view by having a good reference
frame (the cluster stars themselves) from which it is possible to
calibrate systematic effects (e.g., magnitude equation) that
would otherwise
limit the accuracy of proper motions for field stars. In addition,
these data sets provide an opportunity to examine the possibility of
using our model for correcting relative to absolute proper motions
(including secular proper motion) by matching the {\it shape} of the 
observed proper-motion distribution to that predicted by our model.

Completeness limits, areal coverage, and number of stars included in
each survey are shown in Table 3. The bright-magnitude limit was
imposed by the constraint of having good S/N (assumed to be
Poisson-dominated) on the
brightest magnitude bins and also by having enough cluster members to be
able to constrain the magnitude equation on the plates utilized. The number of field stars has been computed
by using the membership probabilities provided in the individual
studies. These memberships were computed using proper motion alone, and proper motion and
spatial probabilities combined. We have adopted the latter, although
the difference in the total counts by using either method is not
significant (see Table 3).


We have performed completeness tests on
both samples by using a modified version of the test presented by
Bienaym\'e et al. (1992). Our nearest-neighbor test gives the number of stars
with closest neighbor at distances
equal to or larger than a given distance. For a uniform distribution of
stars (such as that expected in our field, in the absence of the
clusters), the distribution follows an exponentially decaying function
with the square of the distance. At shorter distances the expected and
observed functions should differ, because of images lost by crowding.
The difference between the two functions evaluated at zero distance
(i.e., including stars at all separations) gives the absolute
completeness of the sample. The absolute completeness is evaluated by
fitting the expected distribution to the observed one at large
distances, where crowding losses do not occur. Our test is more stable
than that used by Bienaym\'e et al. (1992), since the absolute
completeness is calibrated in the regime where losses are not
expected (this regime has been adopted as the distance larger than or
equal to the distance at which the expected cumulative distribution
is independent of the stellar density, equivalent to about 100 arc-sec
in the NGC 188 field and to about 50 arc-sec in the NGC 3680 field).
Figure 1 shows the best $\chi^2$ model fits to the observed
distributions. In order to take into account the presence of the
clusters, membership probabilities have been used to
compute the cumulative numbers. In addition, to take into account the
fact that
we may be searching for the nearest star to a cluster star, we
have excluded all those stars that exceed a certain threshold probability. This
threshold probability has been computed so that the total number of
excluded stars equals the number of cluster stars in the field, as
deduced from the membership probabilities. For NGC 188 we seem to have more stars at smaller separations than expected. This
could be due to some remaining contamination from the cluster which is
difficult to separate from the field, especially at the fainter
magnitudes of the survey. The extant contamination amounts to, {\it at most},
17 \% of the total field star sample. This number is perhaps an
overestimation, because it will imply that about 50\% of the {\it cluster}
sample would be missing, which is rather larger considering the
proper-motion errors quoted by Dinescu et al. (1995). For NGC 3680, we
obtain a much better fit, indicating an incompleteness of 4 to 5\%.
Since both samples will be binned, the quoted incompleteness for NGC
3680, and the possible cluster contamination in the field of NGC 188,
does not alter our conclusions in any significant way.


\subsection{Magnitude and Color counts}

Figure 2 shows the observed {\it vs.} predicted field star magnitude counts
for the fields of NGC 188 and NGC 3680. Figure 3 shows the
color histograms. In all histograms, the observed values represent sums
over the complement of the combined proper-motion and spatial
membership probabilities. The parameters adopted for these runs are 
those indicated
in Table 1. We have taken approximate reddening values towards these two
clusters from the literature. The overall fit to both the color and
magnitude counts for NGC 3680 is remarkable. For NGC 188 the {\it shape}
of the histograms is similar, but the observed number of stars is
larger than expected. This we assume to be due to a larger space
density toward this particular line-of-sight, the derived value being
$0.17 \pm 0.01 stars/pc^3$. This higher value is unlikely to be due
entirely to
cluster contamination which, as we have seen, amounts to less than 17\%, whereas the
observed over-density is around 44\%. Also, the overdensity does not
seem to be a strong function of magnitude, which would be the case if cluster
contamination were its cause. In what follows we have
scaled all our model computations assuming this over-density in
the field of NGC 188, while no scaling has been applied to the model
for the NGC 3680 field.


We have explored the sensitivity of our model predictions to a number
of parameters. For the most sensitive parameters, we have
adopted an iterative scheme that allows us to obtain the best overall
$\chi^2$ fit to the observed magnitude and color counts
simultaneously, while keeping
the least significant parameters fixed to their adopted values. In the
magnitude range of our surveys, the color counts (once convolved with
our photometric errors of $\sim 0.1 \: mag$ in B-V) do not present any
strong signature as a function of magnitude (see also Fig 2b on Lasker
et al. 1987, and Figure 6 on Ojha et al. 1994c). Therefore, 
given the small sample size (particularly in the
case of the NGC 188 field), we have performed our comparisons to the
{\it overall} color distribution in the magnitude range where each
survey is complete (Table 3), rather than breaking the color counts on
different magnitude intervals. Larger samples with smaller
photometric errors ($\leq 0.02 \: mag$ in B-V) would be required to 
cleanly separate field main-sequence from subgiant/giant stars.
Nevertheless, we have
performed some comparisons to our sample broken down by magnitude and
have found that the errors derived for our model parameters (see next
paragraph) from our {\it global} minimization scheme are kept to a
minimum if we use the complete samples, rather than breaking them into
smaller sub-samples according to apparent magnitude or color (this is
also true for our kinematic modeling, see Section 3.3.)

We
find that our model's predictions are not sensitive to the adopted
values for the Solar-galactocentric distance ($R_\odot$) or the
scale-length ($H_R$). This is explained by the fact that the galactic
position of our two fields ($l \sim 123^o$ for the NGC 188 field, $l
\sim 287^o$ for the NGC 3680 field) is such that only a small range in galactocentric distance is
included in the starcounts. For example, at 2 kpc (maximum typical
distance for faintest stars on both surveys), the galactocentric
distance turns out to be 9.6 kpc in the NGC 188 field and 8.2 kpc in 
the NGC 3680 field. Evidently, fields on the meridional plane of the
Galaxy ($l \sim 0^o$ and $l \sim 180^o$) would be more appropriate to
attempt to constrain $R_\odot$ and  $H_R$, as indeed has been done
by Robin et al. (1992) and Ojha et al. (1994a,b). For those parameters that
turn out to be sensitive, we present
in Table 4 their derived values and uncertainties (at the 99\%
confidence interval (CI), or, equivalently, $3\sigma$, using the 
parameter estimation scheme described
by \cite{laet76}). Not surprisingly, the adopted reddening is an
important parameter that affects both the total number of field stars
as well as the color distribution. More interestingly, we have found
that starcounts at intermediate latitudes such
as these allow us to constrain directly the scale-height for subgiants
($M_v < +3.0, \: (B-V)_o > 0.6$), despite previous claims that only
bright ($V< 10$), all-sky, starcounts could reliably constrain this
parameter (\cite{ml83,ba86}).

It should be stated that, since there is
a wide age-span in the Disk for stars which have just left 
the main-sequence ($\sim$ Hyades age at $M_v = +1$ to $\sim$ M67
age at $M_v = +3$, see notes to Table 2), our derived scale-height 
has to be seen as a weighted mean value, involving the star formation rate
in the disk, and the surface density of matter (see, e.g., Equations
(9) and (11) in van der Kruit and Searle 1982, or Equation (16) in
Rohlfs and Wiemer 1982). In the case of giants, \cite{vks82} have
shown that, for a velocity-independent diffusion coefficient
(\cite{wi77}), their scale-height would be
{\it constant}, independent of the details of the star formation
history, which simplifies the interpretation of a single
scale-height for this population. For subgiants, the situation is more
complicated, indeed main-sequence stars in the same absolute magnitude
range exhibit a rapid change of scale-height with $M_v$ (see next
paragraph). However, it is interesting to notice that our
single-number parametrization yields a scale-height which is also
consistent with that for giants. In the case of a velocity-dependent 
coefficient (Wielen and Fuchs 1983, their Equation (27)), the change
with time of the scale-height would be less than 5\% in the
age-interval of disk subgiants, this change would be negligible in
comparison with our statistical uncertainties (see Table 4). As an
extreme case, we have run a model with a scale-height for subgiants
similar to that of main-sequence stars of the same luminosity. The
results are shown in Figures 4 and 5 for the magnitude and color
counts respectively. This assumption about the scale-height provides a
somewhat worse fit to
the magnitude counts, and the discrepancies become more evident in the
color histograms. However, larger samples are clearly desirable to
settle this point.

We have also tested the sensitivity of our derived
parameters, to the {\it form} of the density function away from the
Galactic plane. For this purpose we have considered a
$sech^2(Z/H_z^s)$ function. Continuity of the density and its first derivative
at large densities from the plane (where observations indicate an
exponential decay in stellar density for disk objects) require that

\begin{eqnarray}
H_z^s     & = & 2 \times  H_z \\
\rho^s(0) & = & \frac{\rho(0)}{4}
\end{eqnarray}

where $\rho^s(0)$ and $\rho(0)$ are the densities, at the plane, for
the $sech^2$ and the exponential representations respectively.

Figures 4 and 5 indicate the results of our runs, adopting
the values for $H_z$ (Table 4) and $\rho(0)$ from our Hess-diagram
(see Section 2.1). It can be seen that these particular counts are not
sensitive to our adopted density function, and that, moreover,
continuity at large distances form the plane does provide a
meaningful constraint to explore other plausible density laws on
larger samples.


We are also able to set the overall level for the
scale-height of main-sequence stars ($H_Z(MS)$) in the range 
$+2 \le M_v \le +4$, where the scale-height changes very quickly
with $M_v$. The compilations by Miller and Scalo (1979) and Bahcall and
Soneira (1980), as well as the results from Gilmore and Reid (1983), 
seem to indicate that the scale-height
for main-sequence stars is approximately constant for 
$M_v \le +2$ ($\sim 90 \: pc$)
and for $M_v > +5$ ($\sim 325 \: pc$). Following Bahcall (1986), we
have used a linear interpolation between $M_v = +2$ and $M_v= +5$. We
have considered two extreme cases, taken from the range allowed by
observational data (\cite{gr83}), by assuming a ``lower'' envelope 
and an ``upper''
envelope for $H_Z(MS)$. The lower envelope is described by
a scale-height of 50 pc for early type stars and 300 pc for later type
stars, the upper envelope is described by a scale-height of 120 pc for
early type stars and 400 pc for later type stars. The upper and lower
envelopes are self-similar, in that the slope of the
linear interpolation for $H_Z(MS)$ between the early and late type
stars was kept
constant at the same value adopted by Bahcall (1986), namely, 
$\sim 84 \:  pc/M_v$. The implied values for the scale-height of early
and late type main-sequence stars for both fields is also shown Table 4.

We obtain some marginal sensitivity to the distance of the Sun from
the plane, $Z_\odot$. The results in Table 4 show that the 99\%
confidence interval is extremely wide. We have, therefore, adopted for
$Z_\odot$ the IAU value (\cite{blet59}), namely +7 pc. We should
notice here that Yamagata and
Yoshii (1992) have found a value close to +40 pc for $Z_\odot$ from an 
analysis of starcount data near the north and south galactic poles.
It seems, however, that taking the results for NGC 188 and NGC 3680
together, a value as high as +40 pc is ruled-out by our model.


\subsection{Proper Motions}

The kinematic comparisons take the form of histograms of proper
motions in galactic longitude ($\mu_l$) and galactic latitude
($\mu_b$). Both the observed and the model histograms are fully convolved
with observational errors, so that the comparisons can be carried out
directly. The parameters derived from the magnitude and color counts
described in the previous section were adopted for all the runs
described here. Rather than making the comparisons graphically, we have
looked at statistical descriptors of the proper-motion distribution;
the median proper motion and the proper-motion
dispersion (however, the full shape of the proper-motion histograms
will be used later on for an assesment of the {\it overall} fit to our
model predictions). In order to properly handle outliers, these parameters were
determined from the observed histograms in an iterative way:
Preliminary values for the median and dispersion were
computed, then a window of semi-width three times the proper-motion
dispersion centered on the median was used to recompute the median
and dispersion, until convergence. The derived values for the observed
proper-motion dispersion were also adopted as the window on the model
histograms. In this case, the iterative procedure was the same as for
the observed histogram, except that the window size was kept constant
and equal to six-times the {\it observed} proper-motion dispersion,
centered on the median. We adopted this
procedure so that our comparisons encompass exactly the same
proper-motion interval from the median for both the observed and model
histograms. This is important in view of the different nature of the
extreme outliers on the observed (large errors) and model (velocity ellipsoid wings)
histograms. Therefore, it should be understood that the model proper-motion
dispersions derived here are the values that one would expect
if motions only in the specified proper-motion window are considered.

Figures 6 and 7 show the proper-motion histograms in the field of NGC
188 and NGC 3680 respectively. For these runs, the standard parameters
described in Section 2.2 were used. There is relatively good overall
agreement between the observed and predicted histograms. The
zero-point shift in the histograms is expected, and is due to the
fact that we only have proper motions {\it relative} to
the cluster stars (which have been used to define the reference
frame). Therefore, in principle, we can not use these data to test the
reliability of our predicted proper-motion zero-point. We can, however,
let the zero-point be a free parameter, and compare the {\it shape} of
the observed {\it vs.} predicted histograms. A good match to the shape will
indicate that our model assumptions are appropriate. Even though the
histograms are not completely symmetric (due to the superposition of
velocity ellipsoids with different velocity lags), we have
concentrated mainly on the proper-motion dispersion as a diagnostic of our
model fits. As for the case of the magnitude and color counts, we
have made a number of runs changing the relevant kinematic parameters
successively in order to investigate their effect upon the predicted
median proper motion and the proper-motion dispersion. Table 5
indicates the observed proper-motion dispersions for
these runs as well as their median values. As a general conclusion from these runs one could say
that these particular data are unable to discriminate between a
velocity ellipsoid oriented toward the Galactic axis of rotation from
one oriented towards the Galactic center (q= 0 and q= 1 respectively
in Equation (5), Runs 1 and 2 in Table 5). In addition, the overall
rotation of the disk does not change the predicted velocity
dispersion in a significant way (Runs 3 and 4), a result expected since most stars in
these samples are disk stars which are moving with almost the same
{\it relative} speed. A change of the solar motion affects also mainly
the median proper motion, but not the proper-motion dispersion (Run 5). Still,
the change is quite small, at most 0.6 mas/yr in $\mu_l$ for the NGC
188 field. We have also tried changing the slope of the local rotation
curve, to cover the range of present observational uncertainties,
between $dV_c(R)/dR \sim -15 \: km/s/kpc$ to  $dV_c(R)/dR \sim +5
\: km/s/kpc$ (\cite{kelb86}). Again, only a small change is
seen (Runs 6 and 7).


A common feature that emerges from the runs described above is that the observed and
predicted proper-motion dispersions differ by more than the
observational error of the former. The discrepancy is larger in the
proper-motion dispersion in longitude $\Sigma_{\mu_l}$ than in latitude
$\Sigma_{\mu_b}$. Also, the discrepancy is larger in the NGC 3680 field than in
the NGC 188 field, where the difference between the observed and expected
dispersions reaches $8 \sigma$ of the observed uncertainty. In all
cases, the observed dispersion is larger that the predicted one. Since the observed 
proper-motion dispersion is independent of their 
relative nature, we can attempt to constrain it by comparing to our model predictions directly. Evidently, the most direct way of
increasing the predicted dispersion is to increase either the adopted
proper-motion errors (which are included in the error-convolved
predicted values) or to increase the model velocity
dispersion. In the latter case, the model would allow us to put constraints on
the velocity dispersion for those stars that are mostly represented in our sample. We
have therefore run models with larger errors than those quoted in Dinescu et
al. (1996) and Kozhurina-Platais et al. (1995), and also with increasingly
higher velocity dispersions than
those given in Table 2. It should be emphasized that our adopted velocity dispersion
values have been taken from the compilation by Mihalas and Binney
(1981) and from Ratnatunga et al. (1989). The latter quote
uncertainties of, {\it at most}, 10\% for their derived velocity
dispersions. We have verified with the model that changes in the
adopted velocity dispersion for giants and
subgiants cause a negligible change in the proper-motion dispersion (being bright,
their mean distance is larger than for the fainter stars, and thus
they contribute mainly to the peak of the proper-motion distribution,
while nearby main-sequence stars would contribute mainly to the
proper-motion wings, i.e., they define the proper-motion dispersion). 


Table 5 shows the effect of increasing the quoted errors by $1 \sigma$ of their uncertainty (Run 8). As it can be seen, the dispersion around the
mean proper-motion error (which is a function of apparent magnitude) is small enough that the effect upon the expected values is
minimal. In Runs 9 to 11 we indicate, on the other hand, the effect of increasing the velocity dispersion for
main-sequence stars by 10, 20, and 30\% respectively. In
the field of NGC 188 it seems that an increase close to 20\% would
suffice to fit the observed dispersion in Galactic latitude, and a
30\% increase to fit the dispersion along Galactic longitude. In the
field of NGC 3680 it seems that an increase larger than 30\% would be
required to fit either dispersion. These velocity dispersion increases seem uncomfortably
large considering the relatively small errors for these values quoted
by Ratnatunga et al. (1989), as well as judging from the agreement
between different compilations (\cite{de65,mb81}). Also, it does not
seem possible to fit, simultaneously, the dispersion in both
coordinates and for both lines-of-sight by merely scaling up the local
velocity dispersions. This has prompted us to explore a more interesting dynamical
possibility: Fuchs and Wielen (1987) have computed models for the
change of velocity dispersion as a function of distance from the
Galactic plane under reasonable
gravitational potentials. They find that the velocity dispersion in
all three coordinates must increase with distance from the plane, in
a way almost independent of the assumed potential. We have
incorporated their results in our model, by introducing a linear
increase of velocity dispersion with distance from the plane. Our
adopted values for the slope of the velocity dispersion with
$|Z|$, taken from Fuchs and Wielen (their Figure 4) are given in Table 6. The
results of our runs with this modified functional form for the
velocity dispersion, but retaining the original local velocity
dispersions, are also shown in Table 5 Run 12). We can see a drastic
improvement in the agreement between the observed and predicted
proper-motion dispersions. The effect of a further 10\% increase on the local
velocity dispersions for main-sequence stars (allowed by the
observational uncertainties in these velocity dispersions) is also indicated in
Table 6 (Run 13). This last run provides a better fit to the NGC 3680 data and a
somewhat worse fit to the NGC 188 data than does Run 12. We
conclude therefore that our data shows a clear indication for an increase of
velocity dispersion with distance from the Galactic plane for disk
stars in a way
similar to that predicted by the dynamical models of Fuchs and Wielen
(1987). Kuijken and Gilmore (1989b) have found a similar trend of increasing
$\Sigma_w$ with distance from the Galactic plane from a sample of K
dwarfs at the South Galactic Pole. In this case, radial velocities
could prove only one velocity direction, while, given the Galactic
location of our fields, our proper motions are able to prove,
simultaneously, the trend in $\Sigma_u$ and $\Sigma_w$. The inferred 
value for $d\Sigma_w/dZ$ from Kuijken and Gilmore (1989b, their Figure
13) is 0.0125 km/s/pc in the distance range $|Z|< 1 \, kpc $ , quite
close to our adopted value from Fuchs and Wielen (1987, see Table 6).

In comparing our expected proper-motion distribution to the observed
histogram, there is always the possibility that our velocity
distribution does not provide a good description of the underlying
distribution. Wielen and Fuchs (1983) have shown that the
solution of the Boltzmann equation, including a constant diffusion
term, leads, indeed, to a Schwarzschild distribution for a single 
generation of
stars, where the velocity dispersions increase with age. Assuming a
constant star formation rate, they have evaluated (\cite{fw87}) the
resulting velocity distribution function. They find that the implied
function has a more pronounced peak and more important outer wings
than a Gaussian with the same overall dispersion. This is the natural
result of the superposition of several Schwarzschild-type distributions
with different velocity dispersion. In a recent study, Reid et al.
(1995) have analyzed the kinematics of nearby M dwarfs in the range $8
< M_v < 15$. They find that the overall velocity distribution of these
stars can be quite complex (their Figure 11). However, they
also find that the velocity distribution for the brighter sample ($8 <
M_v < 10$, their Figure 14) {\it is} well reproduced by Gaussians (as
discussed by Reid et al., their fainter sample with $M_v > 12$ may be
subject to a proper-motion bias). Therefore, for our initial models we
have adopted the view that, when properly binned in luminosity (i.e.,
mean age), the velocity distribution of disk stars {\it is} well
represented by a Gaussian function as shown by the available
observations, and as predicted by the theory of stellar diffusion. Of
course, the asymmetric drift will distort these Gaussian functions, but
this effect is fully included in the model (Equations 4 and 5).

In our model, we have modeled the change of velocity dispersion with
age, as a change of velocity dispersion with absolute magnitude (Table
2). In the following paragraph we show that we are indeed able to
reproduce the {\it shape} of the observed proper-motion distribution within
the Poisson noise of our sample, thus providing support to the
correctness of our kinematical assumptions at least at the level
permitted by the current comparisons. This implies that small local
fluctuations in the velocity ellipsoids (due, e.g., to moving groups)
which could make a significant contribution at larger distances from
the plane are not present in the two fields-of-view analyzed here.


As we have pointed out, our motions do not allow us to constrain our
model's absolute proper motions. However, we can still test the
overall fit to the observed proper-motion histogram by using the
best-fit model obtained above. We have implemented a $\chi^2$
procedure to compute the offset between the observed and predicted
proper-motion histograms. Figures 8 and 9 shows the best $\chi^2$ fit to the
observed histogram, using Run 13 (Table 5). The {\it uncertainty} in
the proper-motion offsets at the $1 \sigma$ level is indicated in the
first two rows on Table 7. From
this table we see that our $1 \sigma$ uncertainties in the zero-point
correction are {\it similar} to the
uncertainties in the observed median proper motion (Table 5), which is 
determined by the width of the proper motion distribution (determined
by nature) and the sample size (determined by the observer). This
implies that our fits, for these two samples, are limited not by our 
ability to {\it model} the observed proper-motion histograms correctly
but, rather, by the noise in the
observed histograms themselves. Evidently, a systematic error in our
adopted distances or velocities would be reflected not only by an
offset in the
computed proper motion, but also by the {\it shape} of the proper
motion distribution, which will lead to an uncertainty of
our fits larger than the uncertainty in the median, contrary to our
findings. This opens the prospect for using our model to correct the relative 
proper motions for cluster stars in
these two studies to an absolute reference frame, allowing us to
derive their space velocities. Since the motion of the field stars
relative to the cluster is the same regardless of what reference
frame is used, we have the basic constraint

\begin{equation}
\vec{\mu}_{cl}^a - \vec{\mu}_f^a = \vec{\mu}_{cl}^r - \vec{\mu}_f^r
\end{equation}

where the superindices refer to absolute (a) and relative (r) proper
motion, and the subindices refer to cluster (cl) and field (f). In
proper-motion based membership studies it is customary to use the
cluster stars themselves as the reference frame, due to their smaller
intrinsic proper-motion dispersion. In this case one would have
$\vec{\mu}_{cl}^r = \vec{0}$, and therefore the difference
$\vec{\mu}_f^a -  \vec{\mu}_f^r$ is precisely the absolute motion
of the cluster. This difference is the number whose uncertainty we
have presented in Table 7. In practice, however, a small residual
motion of the cluster is obtained due to the difficulty of selecting,
a priori, only cluster members, so that contamination by
field stars pulls the motion of the cluster to values
different than zero (see e.g. Table 5 in Kozhurina-Platais et al.
(1995) or Table 4 in Dinescu et al. (1996)). For this reason, the
absolute motion of the cluster would be given by $\vec{\mu}_{cl}^a =
\vec{\mu}_f^a + (\vec{\mu}_{cl}^r - \vec{\mu}_f^r)$, where the term
in parentheses is taken directly from the respective proper-motion
studies, and the absolute motion of the field is computed using our model.
The uncertainty of the correction is the propagated
error of the correction derived from $\vec{\mu}_f^a -
\vec{\mu}_f^r$ (which, as we have seen, is essentially the
uncertainty with which one can determine the mean motion of the field
stars, and decreases with the field sample size) plus the error in separating
cluster from field stars.
By using the values for the absolute motion of the field stars quoted in
Table 15 (Run 13), and the relative motion of the cluster relative to field
stars quoted by Dinescu et al. (1996) for NGC 188 and by
Kozhurina-Platais et al. (1995) for NGC 3680, we obtain the absolute
motion of both clusters given in Table 7, which also indicates the
derived tangential velocities, adopting values for the distance modulus
of both clusters from the above papers, indicated in the same table.
The final uncertainty in the derived tangential velocities includes
only uncertainties in the proper motions and not in the distance
modulus, and it is clearly dominated in these two cases by the
uncertainty of locating the (relative) median proper motion of field stars 
rather than the uncertainty of the motion of the cluster relative to the field.
Evidently, the best possible test for our model's ability to reproduce
the observed median absolute proper motion will come from direct comparisons
to absolute proper motions. These comparisons are underway and will be
published elsewhere. However, from the
relatively small variation of the predicted median proper motion seen
in Table 5 for different model assumptions, it is clear that the
corrections derived here are likely to be very close to the value we
would have derived using an absolute reference frame.


\section{Summary}

Comparison of our model to observed starcounts toward the open
clusters NGC 188 and NGC 3680 are successfully used to constrain several Galactic
structure parameters (Table 4). Comparisons to the observed proper motions
indicate that the disk shows a velocity dispersion gradient in a way
similar to that expected from dynamical models, and for which observational
confirmation was awaited (Table 5). In addition, the shape of the observed
proper-motion distribution is well matched, and a correction to absolute
proper motion for both clusters is derived (Table 7). The velocities obtained
for both clusters could be combined with radial velocity studies to
yield orbits for NGC 188 and NGC 3680 under suitable gravitational potentials.

Other lines of sight should be investigated to further validate the
results presented here. In this context, starcounts and proper motions
for field stars in the line-of-sight of intermediate-latitude open clusters
(where reddening is not so severe), provide the best opportunity to
sample the properties of the Galactic disk (within a radius of 2 kpc or
so from the Sun), including the change of kinematical parameters as a
function of position in the Galaxy. In addition, the same analysis would yield
orbits for those open clusters, from which statistical studies of their
orbital parameters and correlations (or lack thereof) with other
parameters (e.g., age and metallicity) could be carried out to
investigate the dynamical and evolutionary status of the Galactic disk.

\acknowledgments

I have greatly benefited from conversations with Dr. T. Girard, and 
Dr. I. Platais, as well as with D. Dinescu and V. 
Kozhurina-Platais. Several suggestions by Dr. S. Majewski and the
anonymous referee(s) have improved the content and scope of this paper. This 
research has been supported in part by NSF and NASA
grants. R. A. M. also acknowledges a travel grant (C-51073) from the 
Chilean Andes Foundation.


\clearpage

\begin{deluxetable}{cccc}
\scriptsize
\tablecolumns{4}
\tablewidth{0pt}
\tablecaption{Starcounts model parameters \label{tab1}}
\tablehead{
\colhead{Parameter} & \colhead{Disk} & \colhead{Thick-Disk} &
\colhead{Halo}
}
\startdata
Luminosity Function & McCuskey (1966) \& Wielen et al. (1983)\tablenotemark{1} & 47 Tuc & M 3 \nl
Hess-Diagram & Robin and Cr\'ez\'e (1986)\tablenotemark{2} & 47 Tuc & M 3 \nl
Scale-length & 3.5 kpc & 3.5 kpc & 2.7 kpc\tablenotemark{3} \nl
Scale-height & 250 pc for giants \& subgiants\tablenotemark{4} & 1.4 kpc & 0.80\tablenotemark{5} \nl
Stellar density & $0.12 \: stars/pc^3$ for $M_v \le 21.5$ 
& \nodata & \nodata \nl
Local normalizations & \nodata & 2\% & 0.125\% \nl
\tablerefs{
(1) see text for details; (2) modified to include subgiants
from the CNS3; (3) this corresponds to the de
Vaucoleurs' half-light radius for an oblate spheroid; (4) 325 pc for
white-dwarfs, function of $M_v$ for main-sequence stars, see text for
details; (5) axial ratio
}
\enddata
\end{deluxetable}

\clearpage

\begin{deluxetable}{cccc}
\tablewidth{0pt}
\tablecaption{Model local velocity dispersions \label{tab2}}
\tablehead{
\colhead{Kinematic group} & \colhead{$\Sigma_u$} & \colhead{$\Sigma_v$} &
\colhead{$\Sigma_w$} \\
\colhead{} & \colhead{km/s} & \colhead{km/s} & \colhead{km/s}
}
\startdata
Giants\tablenotemark{1} & 30 & 20 & 20 \nl
Sub-giants\tablenotemark{2} & 30 & 20 & 20 \nl
Main-sequence & & \nl
$M_v\le +2.2$       & 15 & 10 & 10 \nl
$+2.2< M_v\le +4.8$ & 25 & 15 & 15 \nl
$+4.8< M_v$         & 30 & 20 & 15 \nl
White-dwarfs & 30 & 20 & 20 \nl
\tablerefs{
(1) Giants are stars with $M_v \le -0.7$ and $(B-V)_o >
+1.28$, and $-0.7 < M_v \le +1.0$ and $(B-V)_o> +0.57$; (2) Sub-giants
are stars with $+1.0< M_v \le +3.0$ and $(B-V)_o > +0.57$
}
\enddata
\end{deluxetable}

\clearpage

\begin{deluxetable}{ccc}
\tablewidth{0pt}
\tablecaption{Survey characteristics \label{tab3}}
\tablehead{
\colhead{Parameter} & \colhead{NGC 188 field} & \colhead{NGC 3680 field}
}
\startdata
Areal coverage     & $0.9^oX0.8^o$       & $0.9^oX0.6^o$ \nl
Completeness limit & $12.0 \le V < 15.2$ & $11.0 \le V < 16.8$ \nl
Total sample\tablenotemark{1} & 589 stars & 1626 stars \nl
Number of field stars & & \nl
Using $P_{\mu}$ only\tablenotemark{2} & 430.68 stars & 1563.83 stars \nl
Using $P_{\mu,r}$ \tablenotemark{3}    & 424.33 stars & 1560.00 stars \nl
\tablerefs{
(1) within completeness limit; (2) $P_{\mu}$ is the membership
probability using proper motions exclusively; (3) $P_{\mu,r}$ is the
membership probability using both proper motions and spatial
distribution of cluster stars.
}
\enddata
\end{deluxetable}

\clearpage

\begin{deluxetable}{ccc}
\tablewidth{0pt}
\tablecaption{Derived Galactic structure parameters (all quoted
uncertainties are $3 \sigma$) \label{tab4}}
\tablehead{
\colhead{Parameter} & \colhead{NGC 188 field} & \colhead{NGC 3680 field}
}
\startdata
Reddening\tablenotemark{1} & $0.067 \pm 0.022$ mag & $0.046 \pm 0.017$
mag \nl
$H_z^(Subgiants)$ &  $251 \pm 69$ pc & $249 \pm 38$ pc \nl
$H_z^{young}(Main-seq)$\tablenotemark{2} & $83 \pm 12$ pc  & $78 \pm
8$ pc \nl
$H_z^{old}(Main-seq)$\tablenotemark{2}   & $348 \pm 18$ pc & $340 \pm
11$ pc \nl
$Z_\odot$ & $2 \pm 34$ pc & $-8 \pm 19$ pc \nl
\tablerefs{
(1) value ``at infinity'', scaled down with an exponential-lab of thickness 100
pc; (2) see text for details
}
\enddata
\end{deluxetable}

\clearpage

\begin{deluxetable}{cccccccccc}
\tablecolumns{10}
\tablewidth{0pc}
\scriptsize
\tablecaption{Kinematic values for surveys and models \label{tab5}}
\tablehead{
\colhead{}    &  \multicolumn{4}{c}{NGC 188 field} & \colhead{} &
\multicolumn{4}{c}{NGC 3680 field} \\
\cline{2-5} \cline{7-10} \\
\colhead{Run} & \colhead{$\mu_l^m$} & \colhead{$\mu_b^m$} & 
\colhead{$\sigma_{\mu_l}$} & \colhead{$\sigma_{\mu_b}$} & \colhead{} &
\colhead{$\mu_l^m$} & \colhead{$\mu_b^m$} &
\colhead{$\sigma_{\mu_l}$} & \colhead{$\sigma_{\mu_b}$}
}
\startdata
Observed & $0.47 \pm 0.41$ & $1.38 \pm 0.26$ & $8.25 \pm 0.32$ & $5.10
\pm 0.20$ & & $-0.23 \pm 0.18$ & $-0.08 \pm 0.12$ & $7.12 \pm 0.14$ &
$4.37 \pm 0.09$ \nl
Run 1    & 1.10 & 0.35 & 7.47 & 4.76 & & -7.20 & -1.57 & 5.96 & 3.69\nl
Run 2    & 1.16 & 0.37 & 7.45 & 4.73 & & -7.20 & -1.60 & 5.96 & 3.68\nl
Run 3    & 1.85 & 0.27 & 7.51 & 4.73 & & -6.34 & -1.68 & 6.03 & 3.69\nl
Run 4    & 0.59 & 0.49 & 7.38 & 4.74 & & -8.10 & -1.55 & 5.97 & 3.69\nl
Run 5    & 0.61 & 0.42 & 7.34 & 4.73 & & -6.86 & -1.47 & 5.95 & 3.62\nl
Run 6    & 0.83 & 0.22 & 7.46 & 4.76 & & -7.22 & -1.65 & 5.97 & 3.69\nl
Run 7    & 2.16 & 0.82 & 7.18 & 4.56 & & -7.15 & -1.43 & 6.02 & 3.56\nl
Run 8    & 1.22 & 0.36 & 7.48 & 4.81 & & -7.22 & -1.62 & 6.08 & 3.79\nl
Run 9    & 1.26 & 0.46 & 7.73 & 4.96 & & -7.16 & -1.64 & 6.22 & 3.83\nl
Run 10   & 1.37 & 0.58 & 8.00 & 5.15 & & -7.15 & -1.66 & 6.48 & 3.97\nl
Run 11   & 1.49 & 0.69 & 8.25 & 5.32 & & -7.10 & -1.70 & 6.71 & 4.11\nl
Run 12   & 2.32 & 0.90 & 8.42 & 5.33 & & -7.17 & -1.90 & 6.88 & 4.16\nl
Run 13   & 2.46 & 1.04 & 8.70 & 5.50 & & -7.15 & -1.93 & 7.12 &4.29\nl
\tablerefs{Run 1: Standard values (Table 2), q= 0 (equation (5));
Run2: As Run 1, but q= 1; Run3: $V_c(R)= 180 \: km/s$; Run4: $V_c(R)=
260 \: km/s$; Run5: $\vec{V}_\odot$ from Mihalas and Binney (1981)
rather than the value obtained by Ratnatunga et al. (1989) used in
the previous runs; Run 6: $dV_c(R)/dR \sim +5 \: km/s/kpc$; Run 7:
$dV_c(R)/dR \sim -15 \: km/s/kpc$; Run 8: Errors of proper motion
increased by $1 \sigma$ of their quoted uncertainty; Run 9: 10\%
increase in main-sequence velocity dispersion; Run 10: 20\%
increase in main-sequence velocity dispersion; Run 11: 30\%
increase in main-sequence velocity dispersion; Run 12: Velocity
dispersion wth a $|Z|$ gradient as per Fuchs and Wielen (1987); Run 13:
As run 12 but, in addition, a 10\% increase in main-sequence velocity
dispersion
}
\enddata
\end{deluxetable}

\clearpage

\begin{deluxetable}{cccc}
\tablewidth{0pt}
\tablecaption{Adopted velocity dispersion gradients from the Galactic plane
(from Fuchs and Wielen 1987) \label{tab6}}
\tablehead{
\colhead{Distance from plane} & \colhead{$\frac{d \Sigma_u}{dZ}$} & 
\colhead{$\frac{d \Sigma_v}{dZ}$} & \colhead{$\frac{d \Sigma_w}{dZ}$} \\
\colhead{pc} & \colhead{km/s/pc} & \colhead{km/s/pc} & \colhead{km/s/pc}
}
\startdata
$|Z| < 500$ & $2.72 \, 10^{-2}$ & $1.74 \, 10^{-2}$ & $1.41 \, 10^{-2}$ \nl
$500 \le |Z| < 1000$ & $9.78 \, 10^{-3}$ & $5.44 \, 10^{-3}$ & 
$5.43 \, 10^{-3}$ \nl
$1000 \le |Z|$ & 0.00 & 0.00 & 0.00 \nl
\enddata
\end{deluxetable}

\clearpage

\begin{deluxetable}{ccc}
\tablewidth{0pt}
\tablecaption{Derived tangential motions for the open clusters NGC 188 and NGC 3680 \label{tab7}}
\tablehead{
\colhead{Parameter} & \colhead{NGC 188} & \colhead{NGC 3680}
}
\startdata
$\sigma_{\Delta \mu_l}$ mas/yr & $\pm 0.42$ & $\pm 0.19$ \nl
$\sigma_{\Delta \mu_b}$ mas/yr & $\pm 0.33$ & $\pm 0.11$ \nl
$\mu_{l_{cl}}^a$ mas/yr & $1.17$      & $-7.53$ \nl
$\mu_{b_{cl}}^a$ mas/yr & $-0.26$     & $-1.78$ \nl
m-M mag          & 11.12       & 10.00  \nl
$V_{l_{cl}}$ km/s & $9.3  \pm 3.3$    & $-35.7 \pm 0.9$ \nl
$V_{b_{cl}}$ km/s & $-2.1 \pm 2.7$    & $-8.4  \pm 0.5$ \nl
\enddata
\end{deluxetable}

\clearpage

\figcaption[fig1.ps]{Nearest-neighbor completeness test. Dotted lines
represent the data, dashed lines are the best fits to the
observed distributions in a range of distances where crowding losses
are not expected (100 arc-sec on the NGC 188 field, 50 arc-sec on the
NGC 3680 field). The upper panel is for the NGC 188 field, the lower
is for the NGC 3680 field. \label{fig1}}

\figcaption[fig2.ps]{Starcounts versus apparent V magnitude. Circles
and Poisson error bars denote the data, the dashed line is the model.
The upper panel is for the NGC 188 field, the lower panel is for the
NGC 3680 field. \label{fig2}}

\figcaption[fig3.ps]{Color counts versus reddened B-V color within the
completeness limits in apparent magnitude specified in Table 3. Symbols
as in Figure 2. \label{fig3}}

\figcaption[fig4.ps]{Same as Figure 2, except that a different 
parametrization for the scale-height of subgiants, and a different 
density law, has been used. The continuous line is the best-fit model 
(Table 4), the dashed line is the model with a scale-height for
subgiants as for main-sequence stars, and the dot-dashed line is the
model for a $sech^2$ density law for subgiants. \label{fig4}}

\figcaption[fig5.ps]{Same as Figure 3, symbols as in Figure 4. The
inadequacy of a scale-height for subgiants similar to that of
main-sequence stars is clearly seen in the red wing of the color 
counts. \label{fig5}}

\figcaption[fig6.ps]{Proper-motion histogram of the field of NGC 188
in the magnitude range $12.0 \le V < 15.2$. Symbols as in 
Figure 2. \label{fig6}}

\figcaption[fig7.ps]{Proper-motion histogram of the field of NGC
3680 in the magnitude range $11.0 \le V < 16.8$. Symbols as in Figure
2. \label{fig7}}

\figcaption[fig8.ps]{Same as Figure 6, except that the observed data
have been shifted in proper motion to best match the predicted
distribution. Symbols as in Figure 2. \label{fig8}}

\figcaption[fig9.ps]{Same as Figure 7, except that the observed data
have been shifted in proper motion to best match the predicted
distribution. Symbols as in Figure 2. \label{fig9}}


\clearpage

\begin{thebibliography}{99}

\bibitem[Agekjan and Ogorodnikov 1974]{ao74} Agekjan T. A., and
Ogorodnikov, K.F., 1974, in Highlights of Astronomy, Vol. 3, G. Contopoulos,
Dordrecht: Reidel, 451

\bibitem[Antonuccio-Delogu 1991]{ad91} Antonuccio-Delogu, V., 1991,
\aap, 247, 45

\bibitem[Bahcall and Soneira 1980]{bs80} Bahcall, J. N., and Soneira,
R. M., 1980, \apj, 73, 110

\bibitem[Bahcall 1986]{ba86} Bahcall, J.N., 1986, Ann. Rev. Astron.
and Astrophys., 24, 557

\bibitem[Bahcall et al. 1987]{bcr87} Bahcall, J. N., Casertano, S., and
Ratnatunga, K. U., 1987, \aj, 320, 515

\bibitem[Bienaym\'e et al. 1992]{biet92} Bienaym\'e, O. Mohan, V.,
Cr\'ez\'e, M., Consider\`e, S., and Robin, A. C., 1992, \aap, 253, 389

\bibitem[Binney and Tremaine 1987]{bt87} Binney, J., and Tremaine, S.,
1987, Galactic Dynamics, New Jersey: Princeton

\bibitem[Blaauw et al. 1959]{blet59} Blaauw, A., Gunn, C. S., Pawsey,
J. L., and Westerhout, G., 1959, M. N. R. A. S., 119, 422

\bibitem[Colles et al. 1991]{coet91} Colles, M., Ellis, R.S., Taylor,
K., and Shaw, G., 1991, M.N.R.A.S., 253, 686

\bibitem[Delhaye 1965]{de65} Delhaye, J., 1965, in Galactic Structure,
Stars and Stellar Systems, Vol. 5, edited by Adriaan Blaauw and
Maarten Schmidt (The University of Chicago, Chicago), page 61

\bibitem[Dinescu et al. 1996]{di95} Dinescu, D. I., Girard, T. M., van
Altena, W. F., Yang, T.-G., and Lee, Y.-W., 1996, \aj, 111, 1205

\bibitem[Fricke 1952]{fr52} Fricke, W., 1952, Astr. Nach., 280, 193

\bibitem[Fuchs and Wielen 1987]{fw87} Fuchs, B., and Wielen, R., 1987,
The Galaxy, NATO-ASI series, G. Gilmore, B. Carswell, Dordrecht:
Reidel, 375

\bibitem[Gilmore 1984]{gi84} Gilmore, G., 1984, M.N.R.A.S., 207, 223

\bibitem[Gilmore and Reid 1983]{gr83} Gilmore, G., and Reid, N., 1983,
M.N.R.A.S., 202, 1025

\bibitem[Gilmore et al. 1989]{gwk89} Gilmore, G., Wyse, R. F. G., and
Kuijken, K., 1989, Ann. Rev. Astron. Astrophys., 27, 555

\bibitem[Hoffleit 1982]{ho82} Hoffleit, D., 1982, The Bright Star Catalogue,
4th ed, New Haven, Yale University Observatory

\bibitem[Humphreys 1993]{rh93} Humphreys, R., 1993, Workshop on
Databases for Galactic Structure, A. G. Davis Philip, B. Hauck, A. R.
Upgren, Schenectady: L. Davis Press, 87

\bibitem[Jahreiss and Gliese 1993]{jg93} Jahreiss H., and Gliese W.,
1993, Workshop on Databases for Galactic Structure, A. G. Davis
Philip, B. Hauck, and A. R. Upgren, Schenectady: L. Davis Press, 53

\bibitem[Jahreiss 1993]{ja93} Jahreiss H., 1993, private communication

\bibitem[Kent et al. 1991]{keet91} Kent, S. M., Dame, T. M., and
Fazio, G., 1991, \apj, 378, 131

\bibitem[Kerr and Lynden-Bell 1986]{kelb86} Kerr, F. J., and
Lynden-Bell, D., 1986, M. N. R. A. S., 221, 1023

\bibitem[King 1990]{ki90} King, I. R., 1990, in The Milky Way as a
Galaxy, Saas-fee Advanced course No. 19, Roland Buser, Ivan R. King,
California: University Science, page 117

\bibitem[Kirkpatrick et al. 1994]{kiet94} Kirkpatrick, J. D., McGraw,
J. T., Hess, T. R., Liebert, J., McCarthy, D. W., 1994, Ap. J. 
Suppl., 94, 749

\bibitem[Klemola et al. 1987]{klet87} Klemola, A. R., Jones, B. F.,
and Hanson, R. B., 1987, \aj, 94, 501

\bibitem[Kozhurina-Platais et al. 1995]{kp95} Kozhurina-Platais, V.,
Girard, T. M., Platais, I., van Altena, W. F., Ianna, P. A., and
Cannon, R. D., 1995, \aj, 109, 672

\bibitem[Kuijken and Gilmore 1989a]{kg89a} Kuijken, K., and Gilmore, G.,
1989a, M.N.R.A.S., 239, 571

\bibitem[Kuijken and Gilmore 1989b]{kg89b} Kuijken, K., and Gilmore, G.,
1989b, M.N.R.A.S., 239, 605

\bibitem[Lewis and Freeman 1989]{lf89} Lewis, J. R., and Freeman, K.
C., 1989, \aj, 97, 139

\bibitem[Lampton et al. 1976]{laet76} Lampton, M., Margon, B., and
Bowyer S., 1976, \apj, 208, 177

\bibitem[Lasker et al. 1987]{laset87} Lasker, B. M., Garnavich, P. M.,
and Reynolds, A. P., 1987, \apj, 320, 502

\bibitem[Maciel and Dutra 1992]{md92} Maciel, W. J., and Dutra, C. M.,
1992, \aap, 262, 271

\bibitem[Majewski 1993]{ma93} Majewski, S. R., 1993, Ann. Rev. Astron.
Astrophys, 31, 575

\bibitem[McCuskey 1966]{mc66} McCuskey, S. W., 1966, in Vistas in Astronomy,
Vol. 7, A. Beer, Oxford: Pergamon, 141

\bibitem[McLaughlin 1983]{ml83} McLaughlin, S. S., 1983, \aj, 88, 1633

\bibitem[Mihalas and Binney 1981]{mb81} Mihalas, D., and Binney, J.,
1981, Galactic Astronomy,New York: Freeman

\bibitem[Miller and Scalo 1979]{ms79} Miller, G. E., and Scalo, J. M.,
1979, \apj, 41, 513

\bibitem[Monet 1988]{mon88} Monet, D. G., 1988, Ann. Rev. Astron
Astrophysics, 26, 413

\bibitem[Ojha et al. 1994a]{ojet94a} Ojha, D. K., Bienaym\'e, O., and
Robin, A. C., 1994a, in Astronomy from Wide-Field Imaging, IAU Symposium
161, H. T. MacGillivray, E. B. Thomson, B. M. Lasker, I. N. Reid, D.
F. Malin, R. M. West, and H. Lorenz, Dordrecht: Kluwer, 447

\bibitem[Ojha et al. 1994b]{ojet94b} Ojha, D. K., Bienaym\'e, O.,
Robin, A. C., and Mohan, V., 1994b, \aap, 284, 810

\bibitem[Ojha et al. 1994c]{ojet94c} Ojha, D. K., Bienaym\'e, O.,
Robin, A. C., and Mohan, V., 1994c, \aap, 290, 771

\bibitem[Ojha et al. 1996]{ojet96} Ojha, D. K., Bienaym\'e, O.,
Robin, A. C., Cr\'ez\'e, M., and Mohan, V., 1996, to appear in
\aap

\bibitem[Ratnatunga et al. 1989]{rbc89} Ratnatunga, K. U., Bahcall, J.
N., and Casertano, S., 1989, \apj, 339, 106

\bibitem[Reid and Majewski 1993]{rema93} Reid, N., and Majewski, S. R.,
1993, \apj, 409, 635

\bibitem[Reid et al. 1995]{reet95} Reid, I. N., Hawley, S. L., and
Gizis, J. E., 1995, \aj, 110, 1838

\bibitem[Robin and Cr\'ez\'e]{rocr86} Robin, A., Cr\'ez\'e, M., 1986,
\aap, 151, 71

\bibitem[Robin and Oblak 1987]{roob87} Robin, A. C., and Oblak, E., 1987,
Evolution of Galaxies, 10th IAU European Regional Meeting, Vol. 4, J.
Palous, Ondrejov: Astronomical Institute of the Czechoslovak Academy of
Sciences, 323

\bibitem[Robin et al. 1992]{rocr92} Robin, A.C., Cr\'ez\'e, M., and
Mohan, V., 1992, \aap, 265, 32

\bibitem[Robin et al. 1996]{roet96} Robin, A. C., Haywood, M.,
Cr\'ez\'e, M., Ojha, D. K., and Bienaym\'e, O., 1996, to appear in
\aap

\bibitem[Rohlfs and Wiemer 1982]{rw82} Rohlfs, K., and Wiemer H.-J.,
1982, \aap, 112, 116

\bibitem[Schwarzschild 1907]{sc07} Schwarzschild, K., 1907, Akad.
Wissenschaft Gotingen Nach., 614

\bibitem[Schwarzschild 1907]{sc08} Schwarzschild, K., 1908, Akad.
Wissenschaft Gotingen Nach., 191

\bibitem[Shu 1969]{s69} Shu, F., 1969, \apj, 158, 505

\bibitem[Sommer-Larsen 1991]{sl91} Sommer-Larsen, J., 1991, M.N.R.A.S.,
249, 368

\bibitem[Soubiran 1992]{so92} Soubiran, C., 1992, \aap, 274, 181

\bibitem[Spitzer 1942]{s42} Spitzer, L., 1942, \apj, 95, 329

\bibitem[Trumpler and Weaver 1962]{tw62} Trumpler, R. J., and Weaver,
H. F., 1962, Statistical Astronomy, New York: Dover

\bibitem[van Altena et al. 1994]{vaet94} van Altena, W. F., Platais,
I., Girard, T. M., and Lopez, C. E., 1994, in Galactic and Solar
System Optical Astrometry, L. V. Morrison and G. F. Gilmore,
Cambridge: Cambridge Univ. Press, 26

\bibitem[van der Kruit and Searle 1981]{vks81} van der Kruit, P. C.,
and Searle, L., 1981, \aap, 105, 115

\bibitem[van der Kruit and Searle 1982]{vks82} van der Kruit, P. C.,
and Searle, L., 1982, \aap, 110, 61

\bibitem[Villumsen and Binney 1985]{vb85} Villumsen, J. V., and
Binney, J., 1985, \apj, 295, 388

\bibitem[von Hippel and Bothun 1993]{hb93} von Hippel, T., Bothun, G.
D., 1993, \apj, 407, 115

\bibitem[Wielen and Fuchs 1983]{wifu83} Wielen, R., and Fuchs, B.,
1983, in Kinematics, Dynamics, and Structure of the Milky Way, W. L.
H. Shuter, Dordrecht: Reidel, 81

\bibitem[Wielen 1977]{wi77} Wielen, R., 1977, \aap, 60, 263

\bibitem[Wielen et al. 1983]{wi83} Wielen, R., Jahreiss, H., and
Kruger, R., 1983, in The nearby stars and the Stellar Luminosity
Function, IAU Colloq. 76, A. G. Davis Philip and A. R.
Upgren, Schenectady: L. Davis Press, 163

\bibitem[Woolley et al. 1970]{wo70} Woolley, R., Epps, E. A.,
Penston, M. J., and Pocock, B., 1970, Roy. Obs. Ann., 5

\bibitem[Yamagata and Yoshii 1992]{yayo92} Yamagata, T., and Yoshii,
Y., 1992, \aj, 103, 117

\bibitem[Zjilstra and Pottasch 1991]{zp91} Zjilstra, A. A., and
Pottasch, S. R., 1991, \aap, 243, 478

\end{thebibliography}
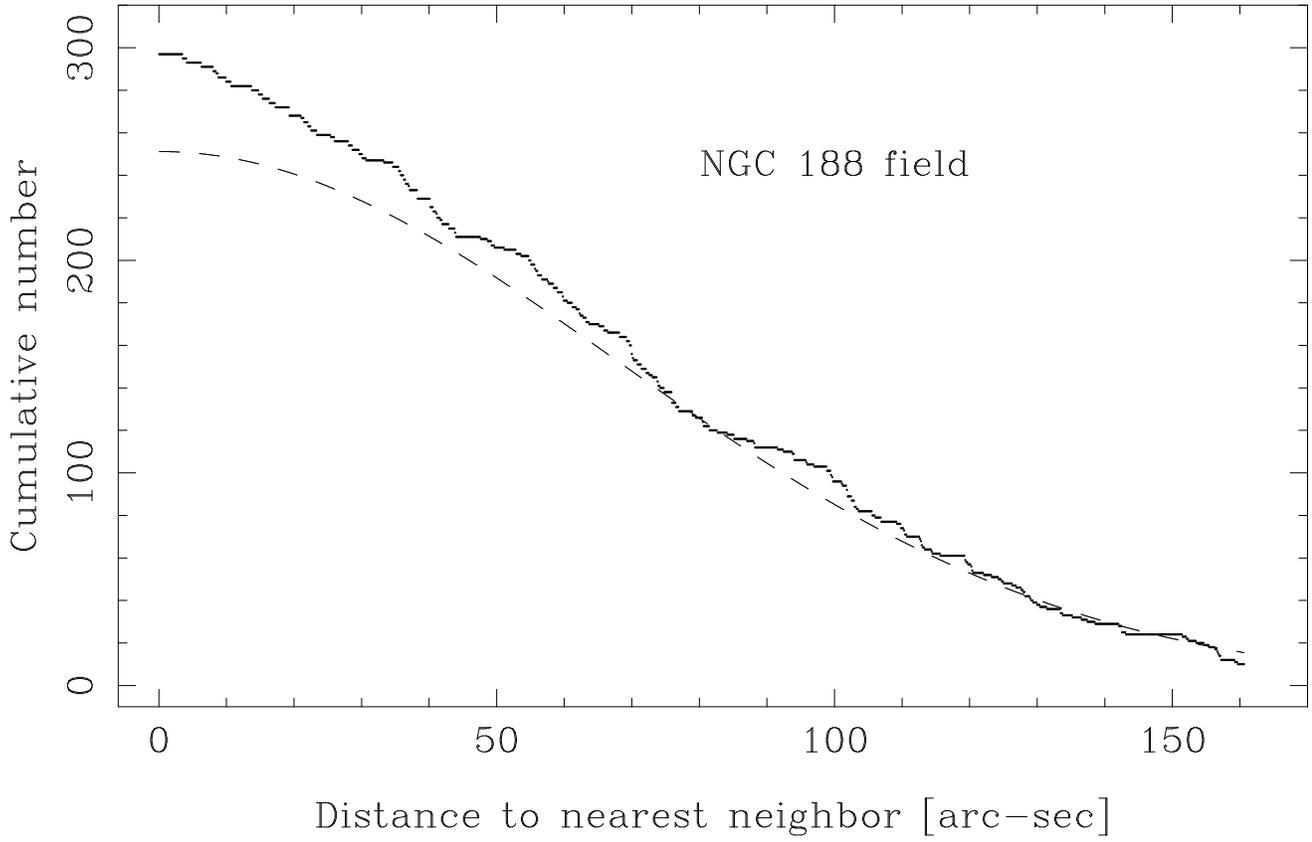
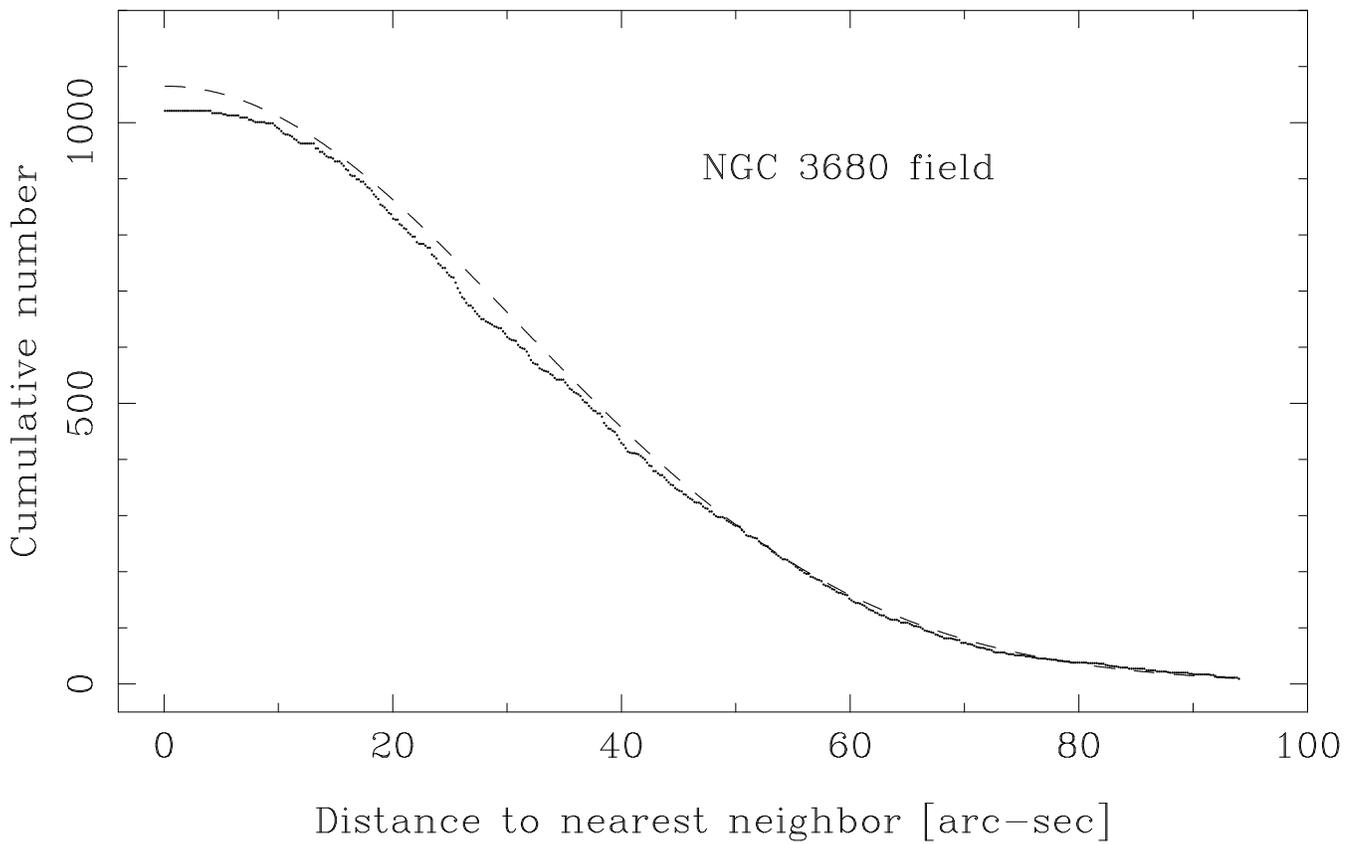
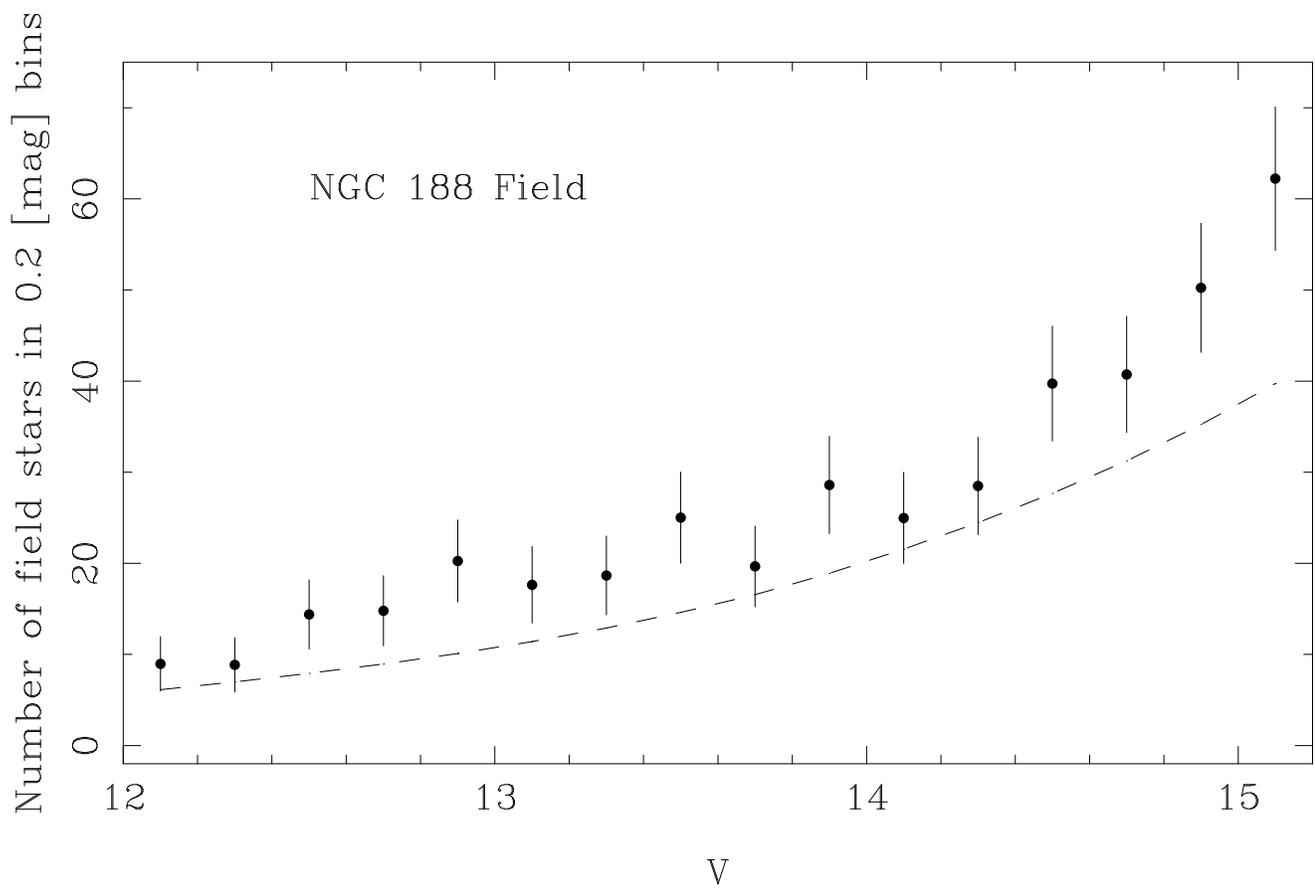
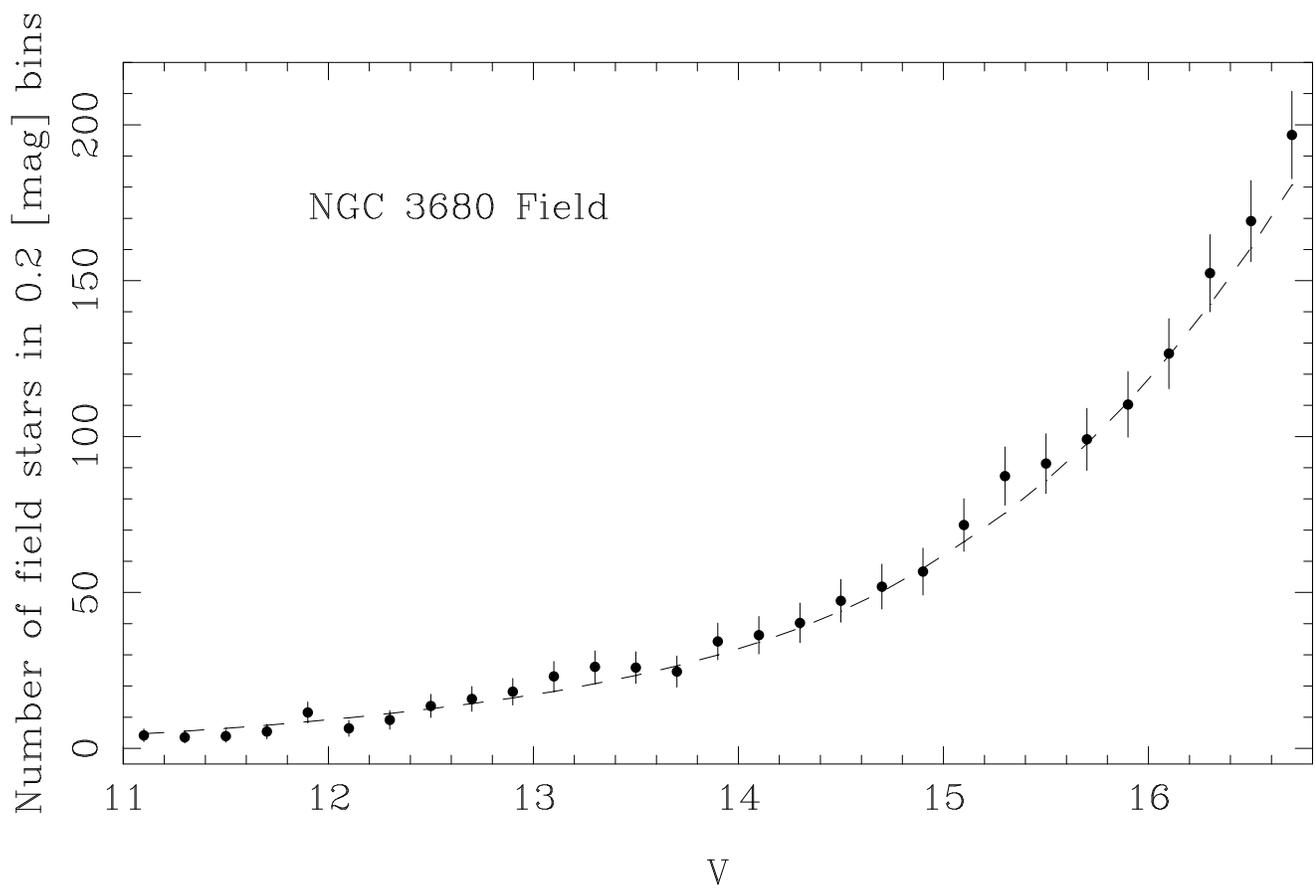
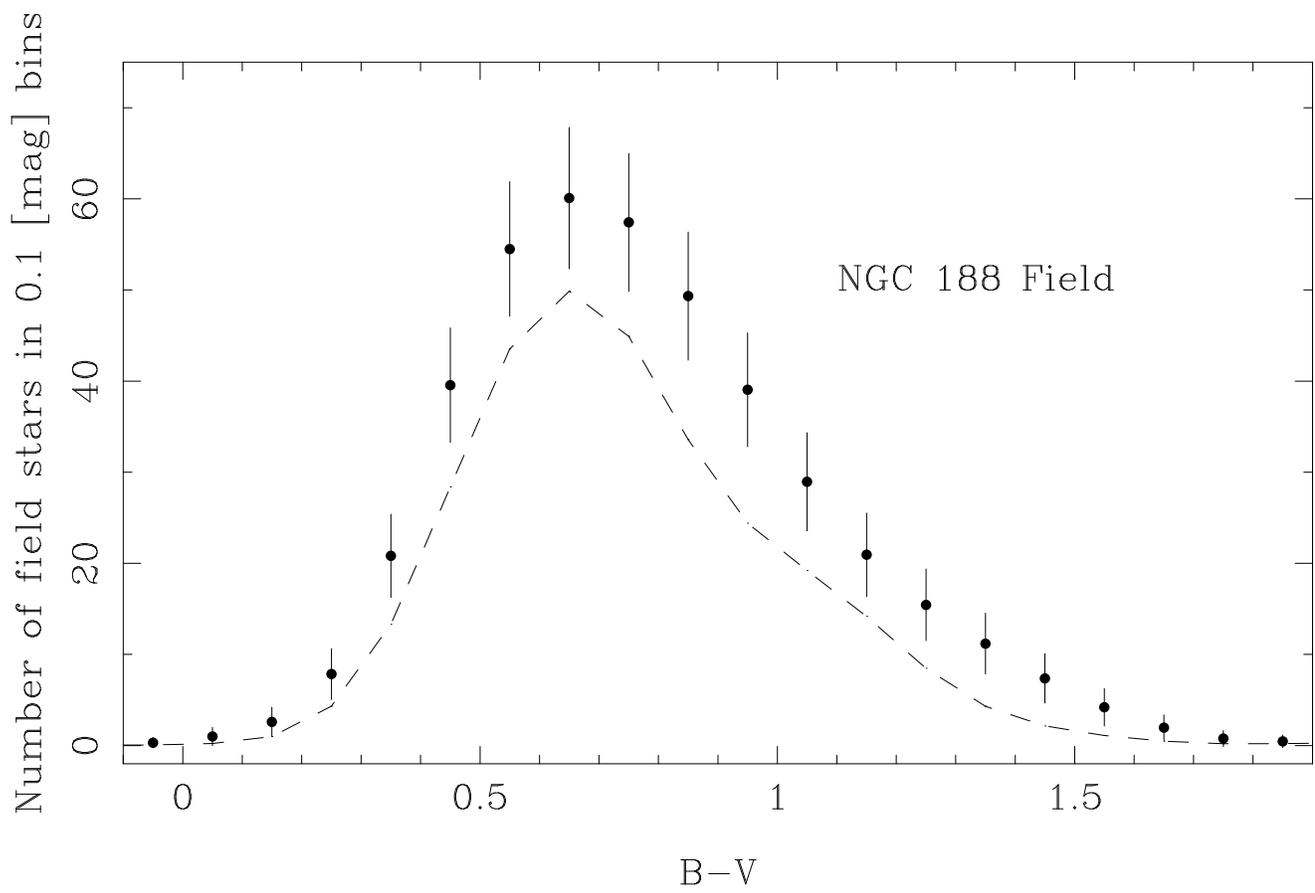
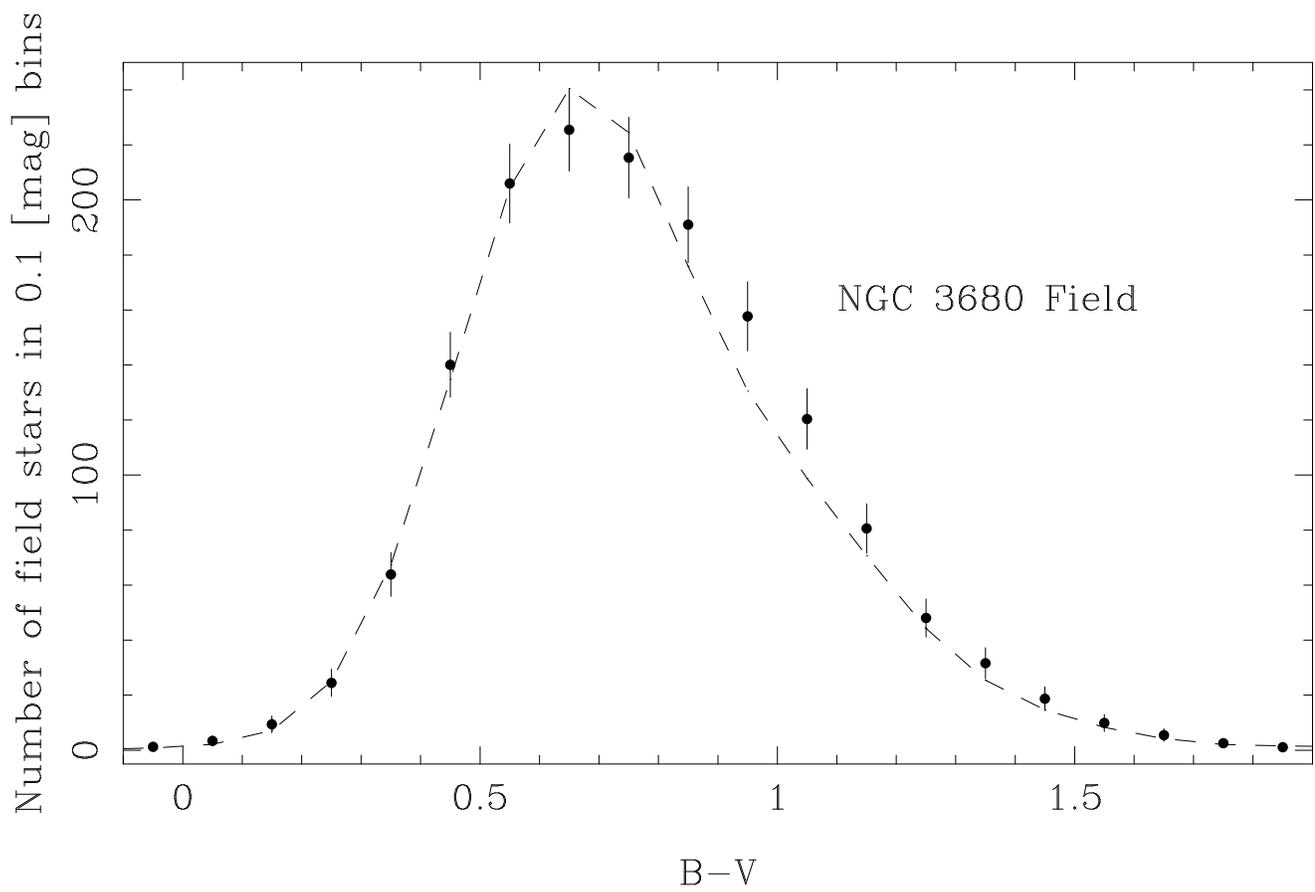
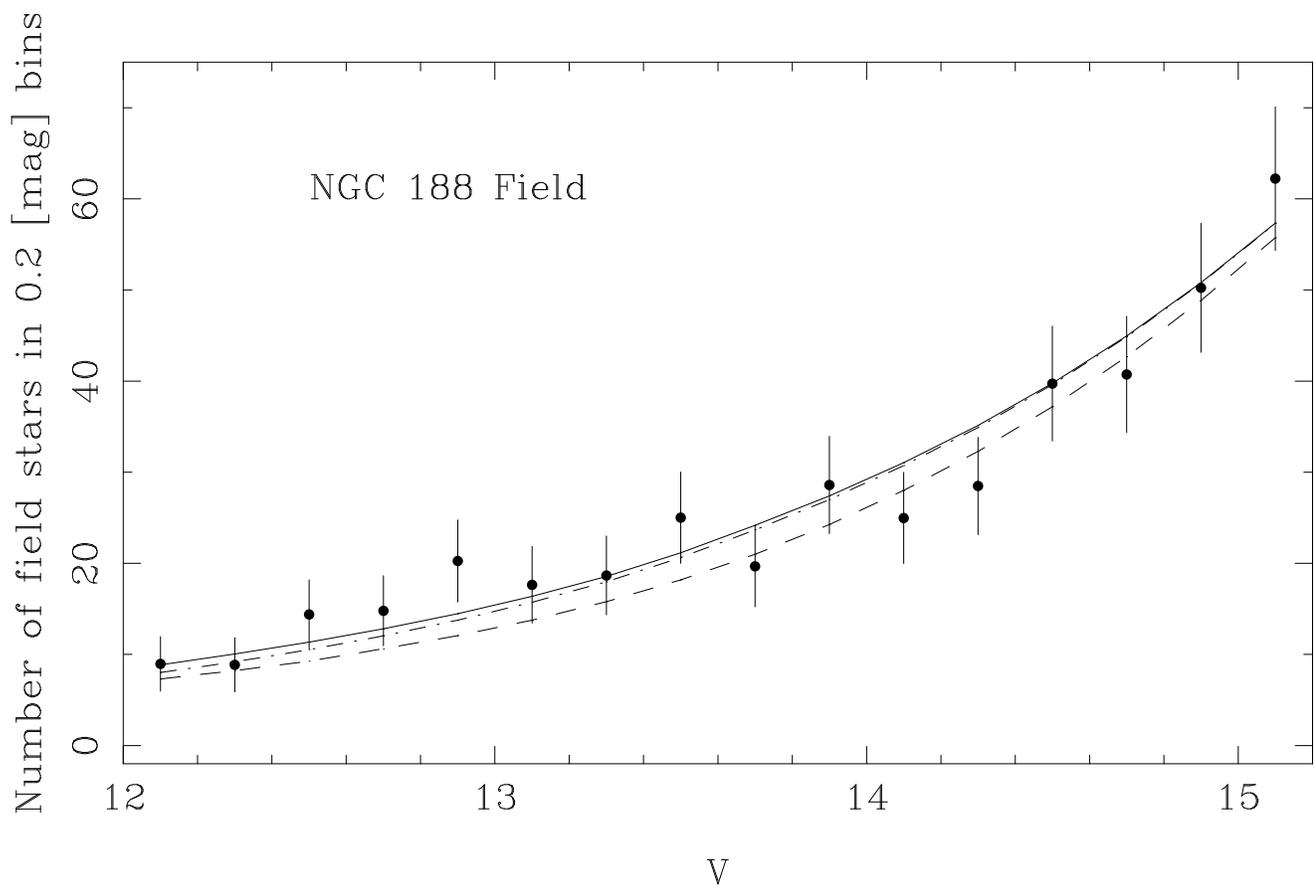
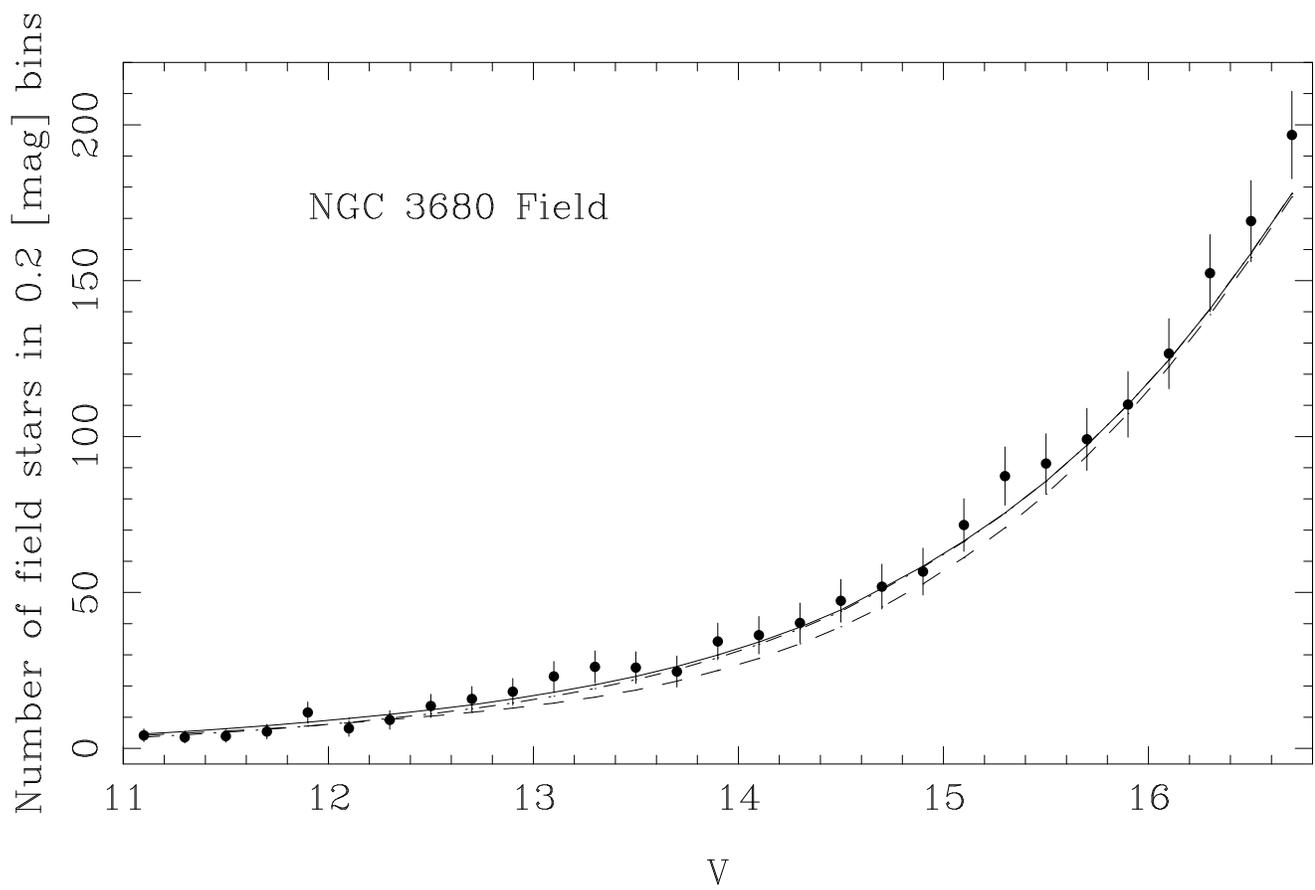
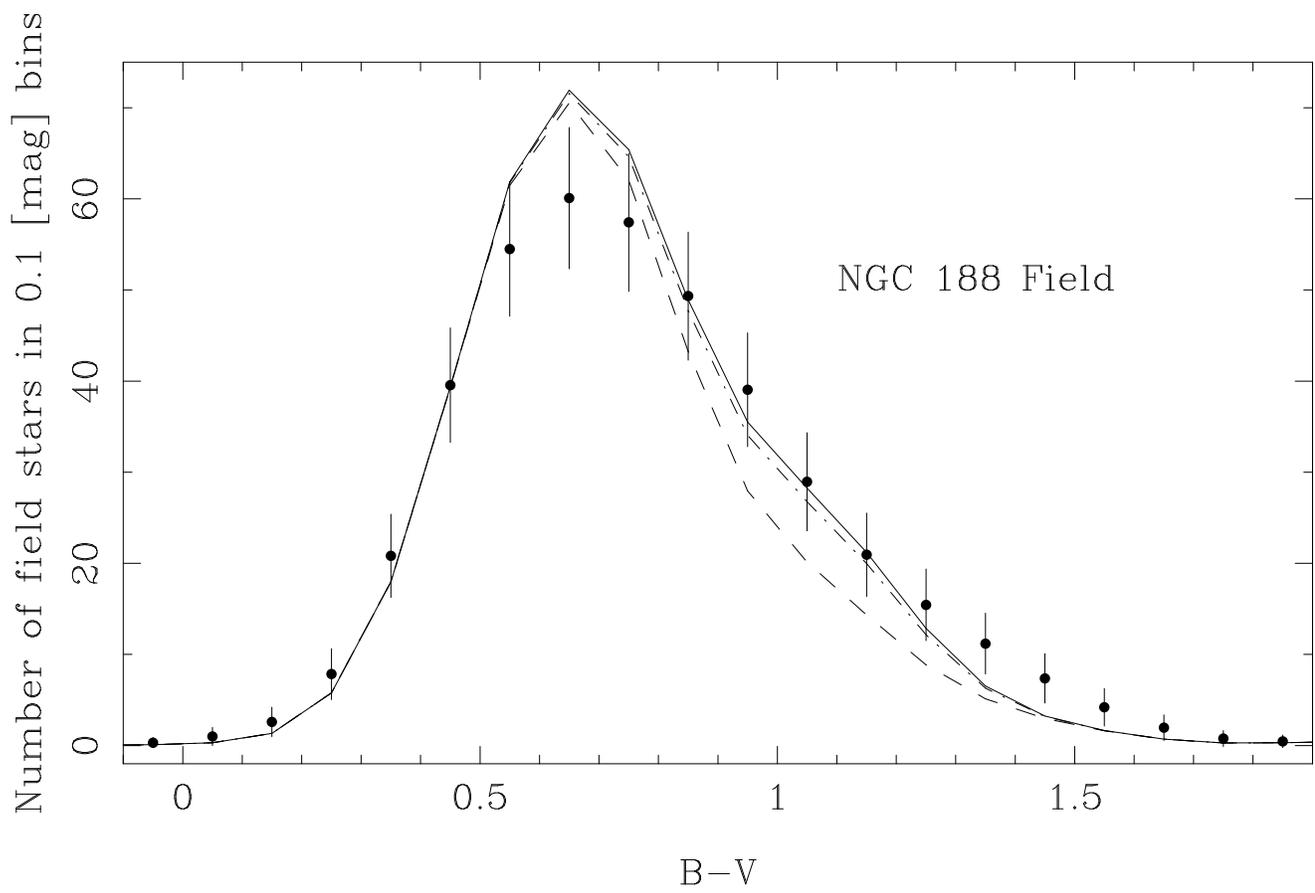
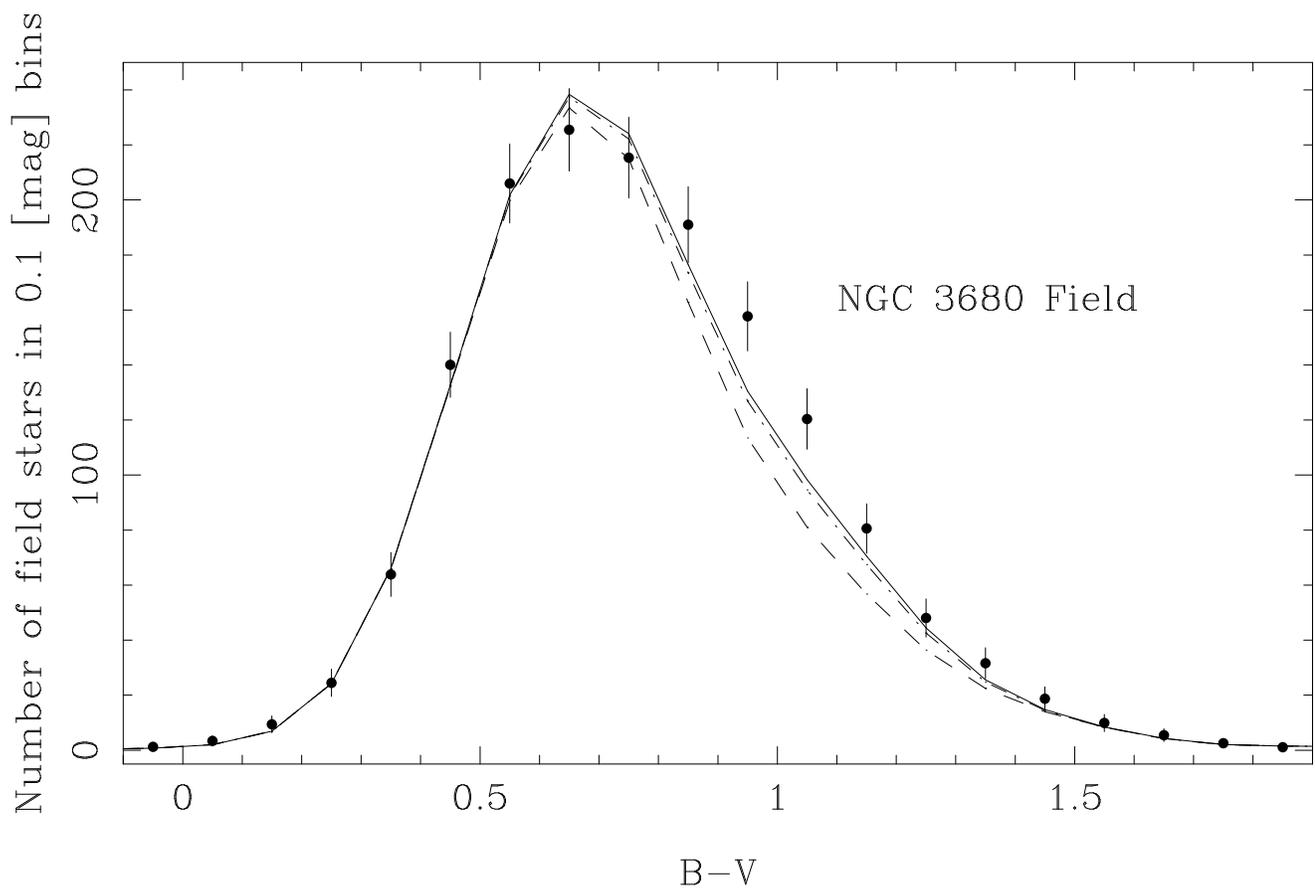
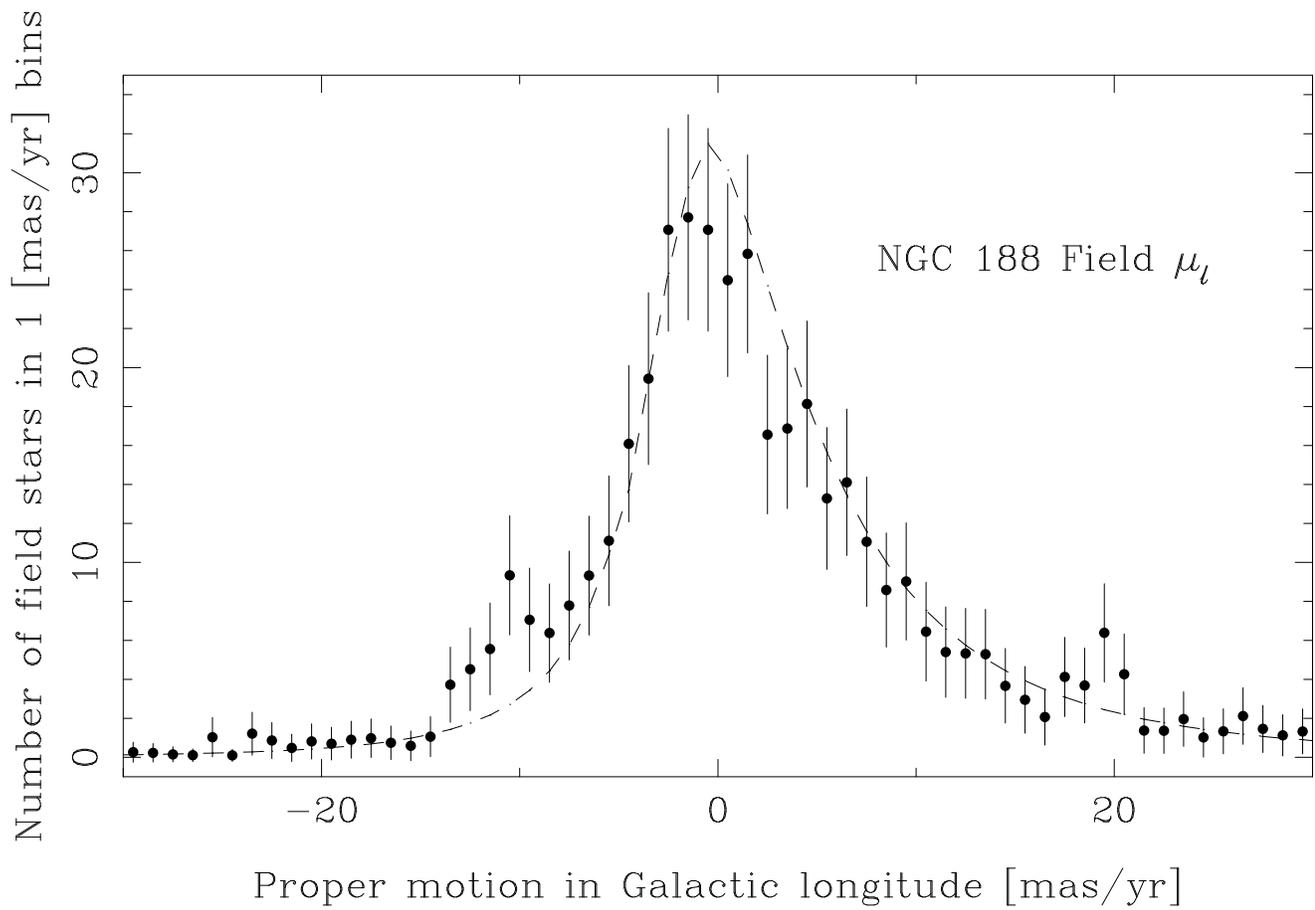
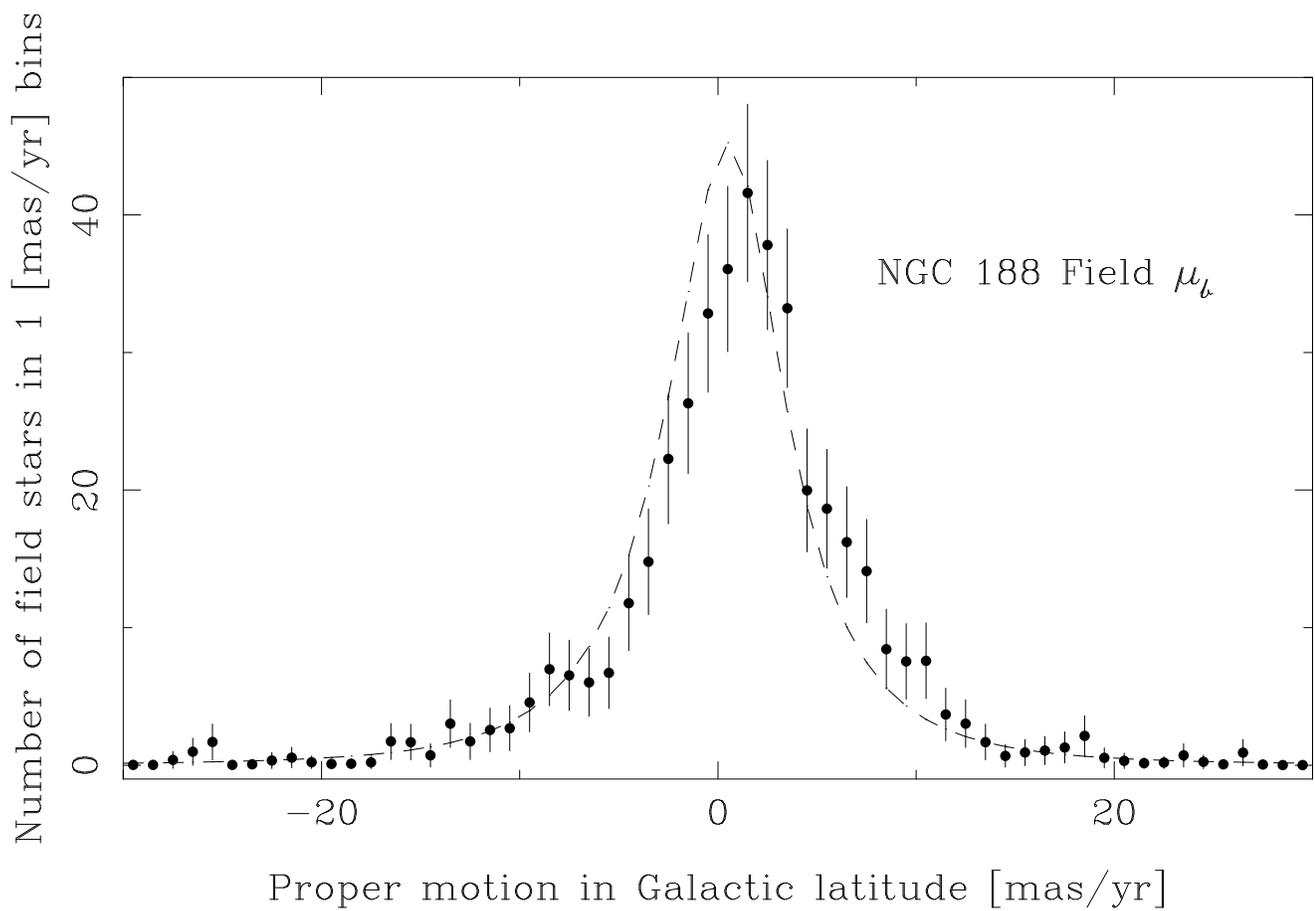
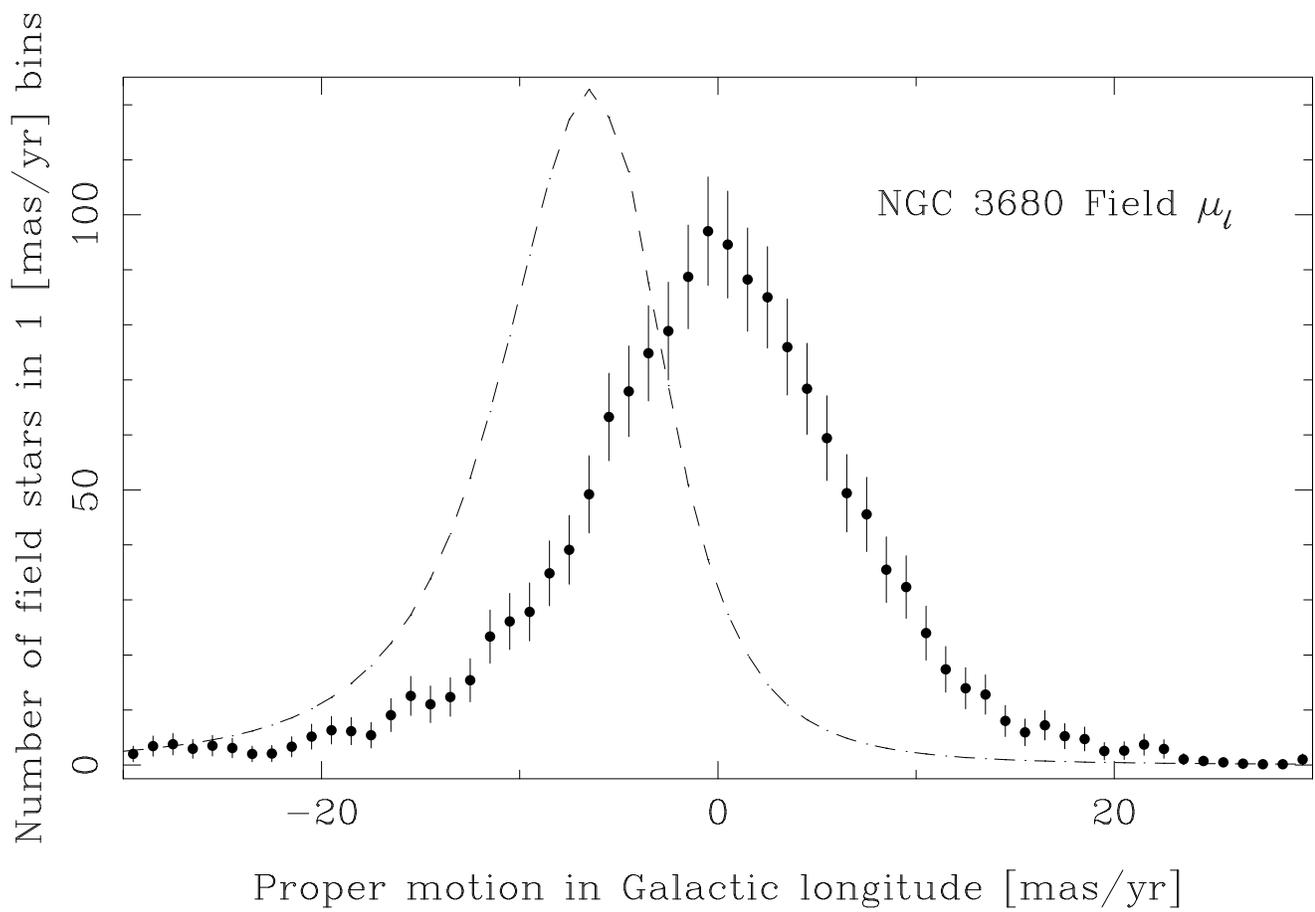
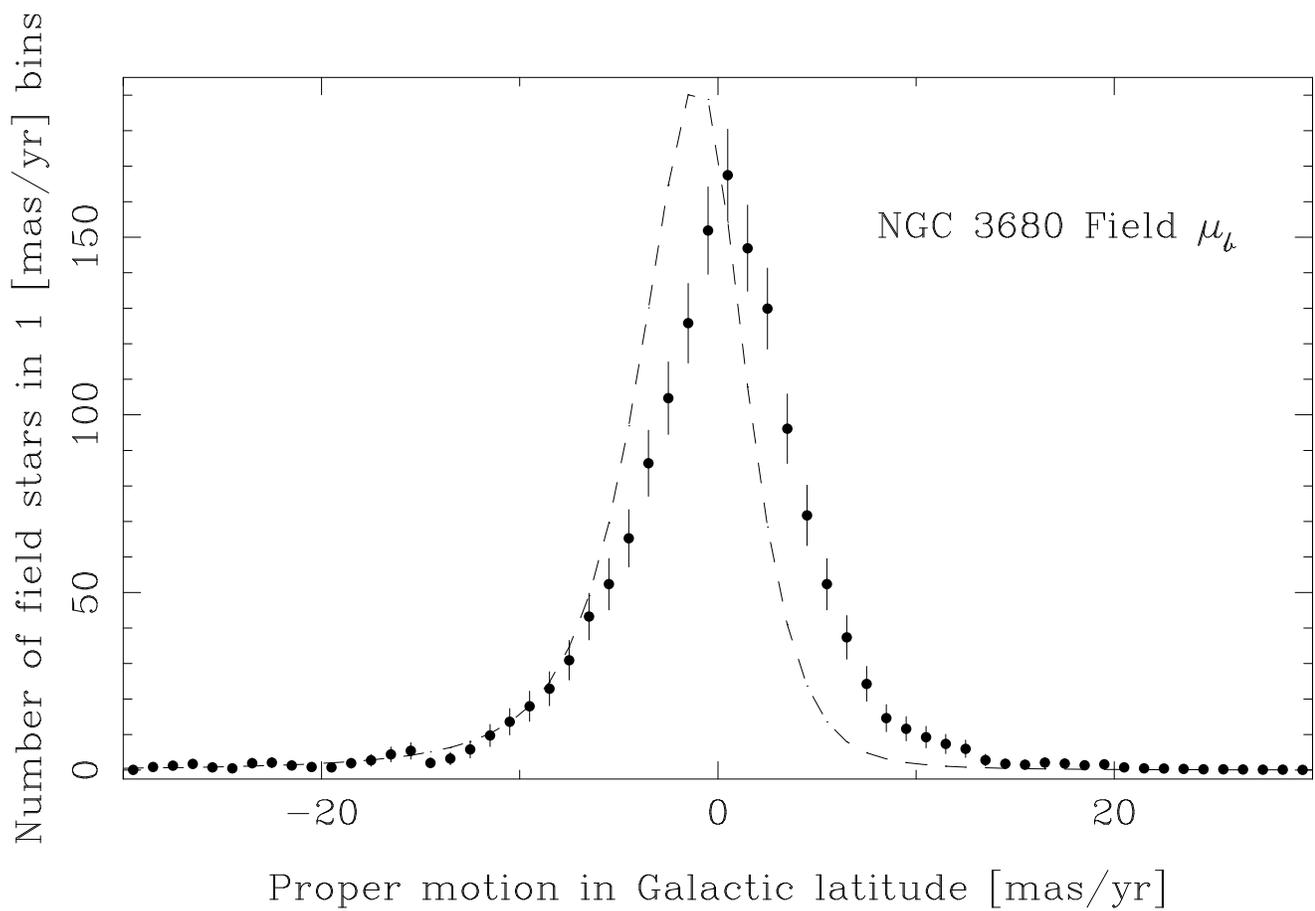
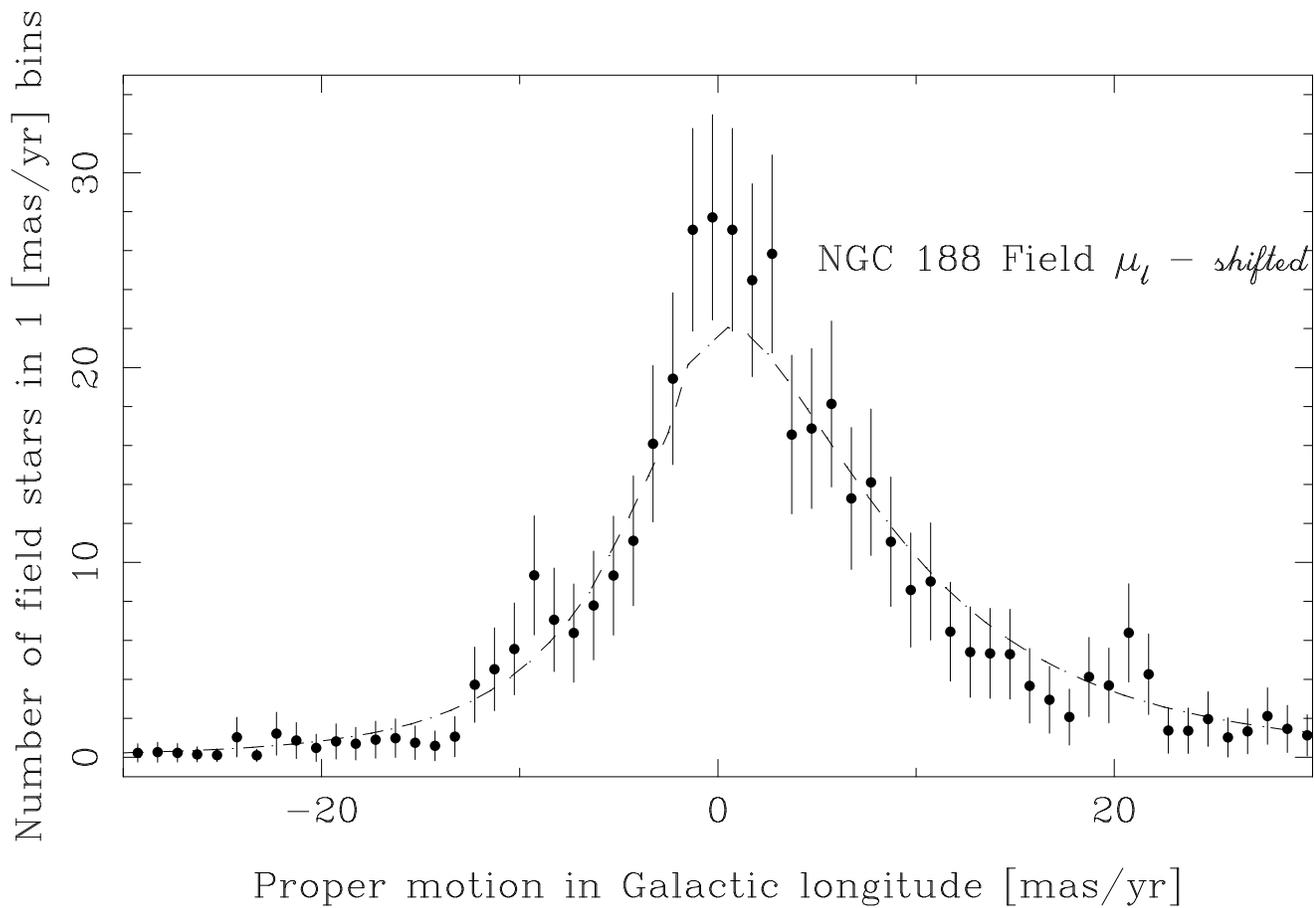
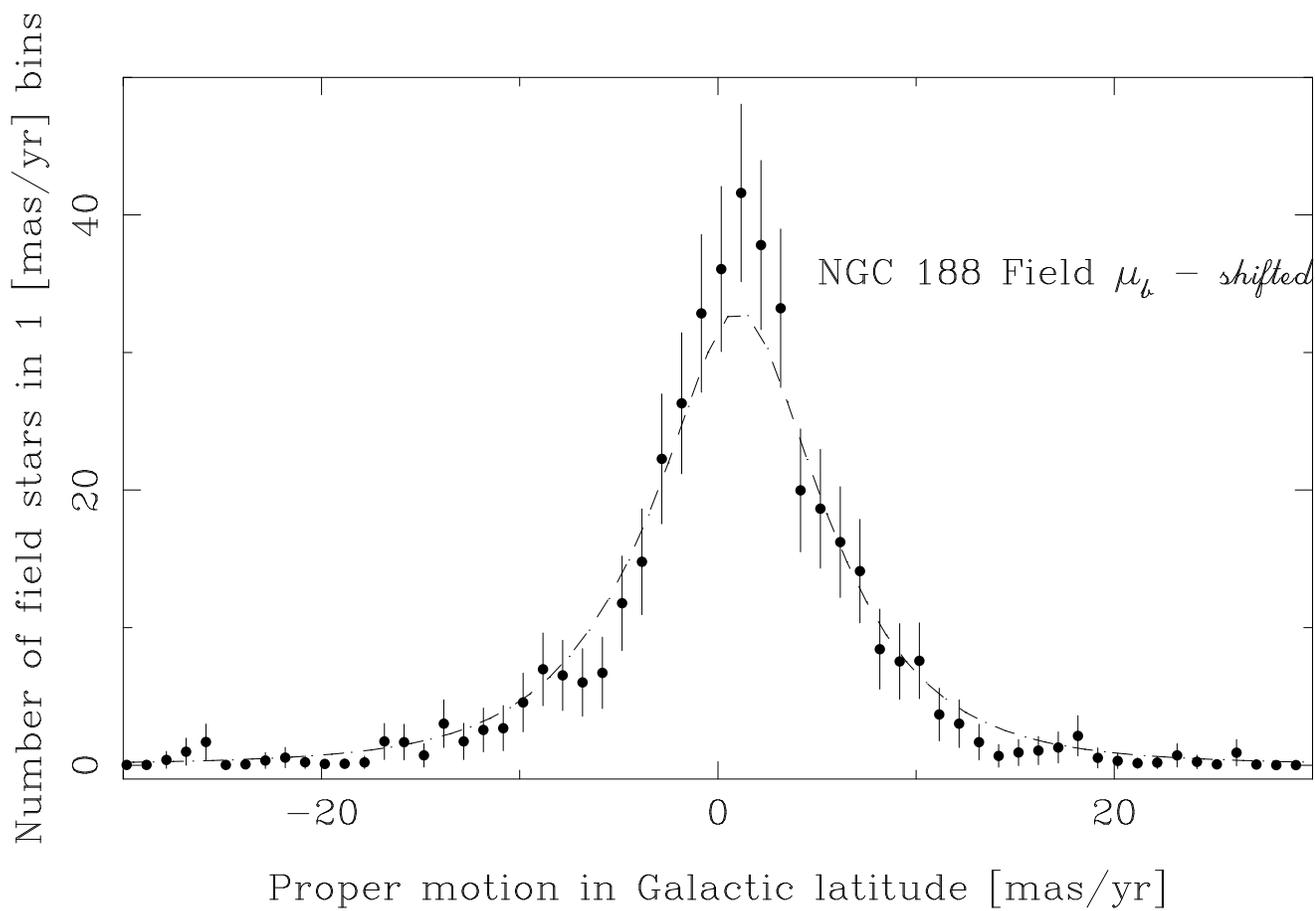
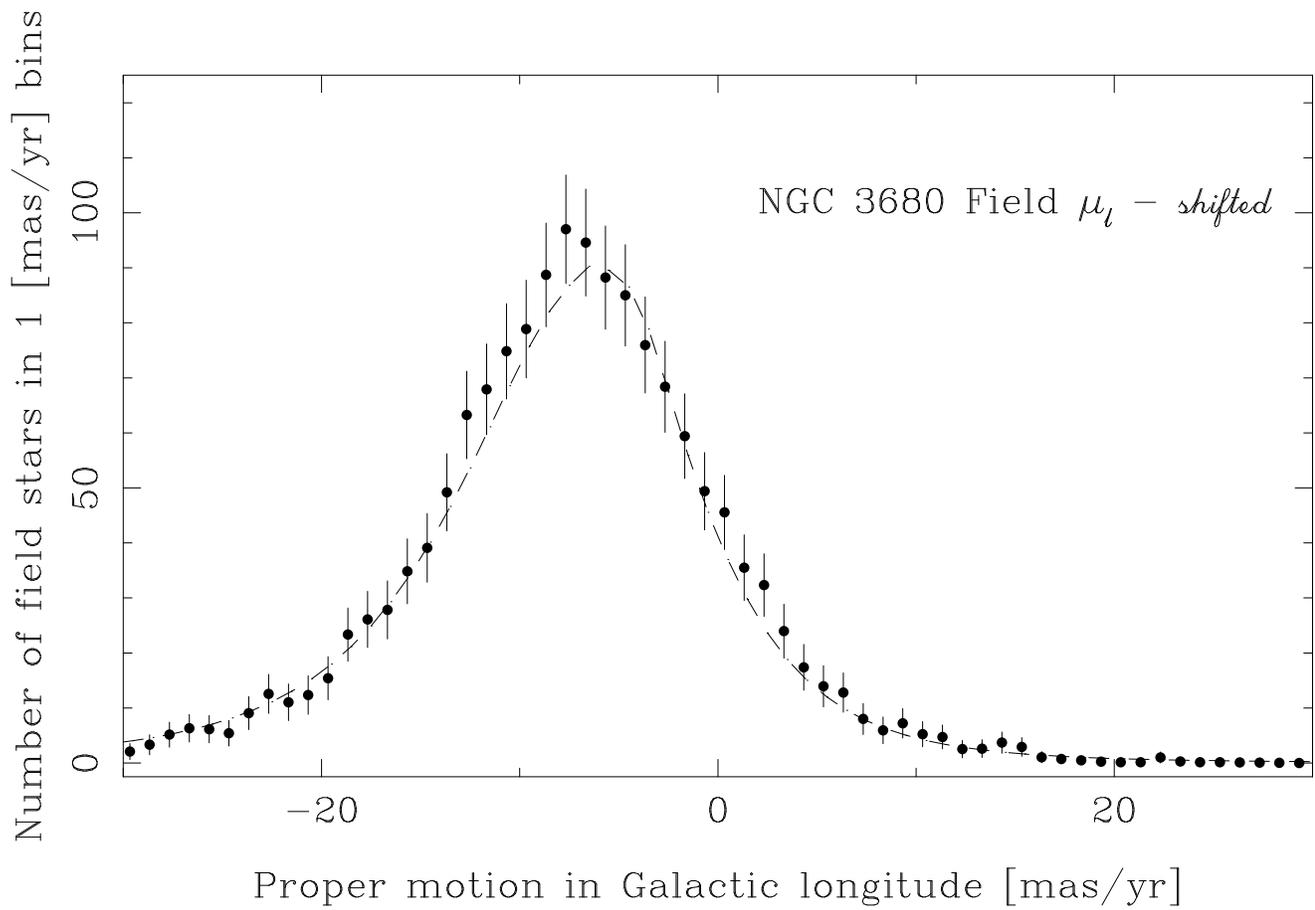
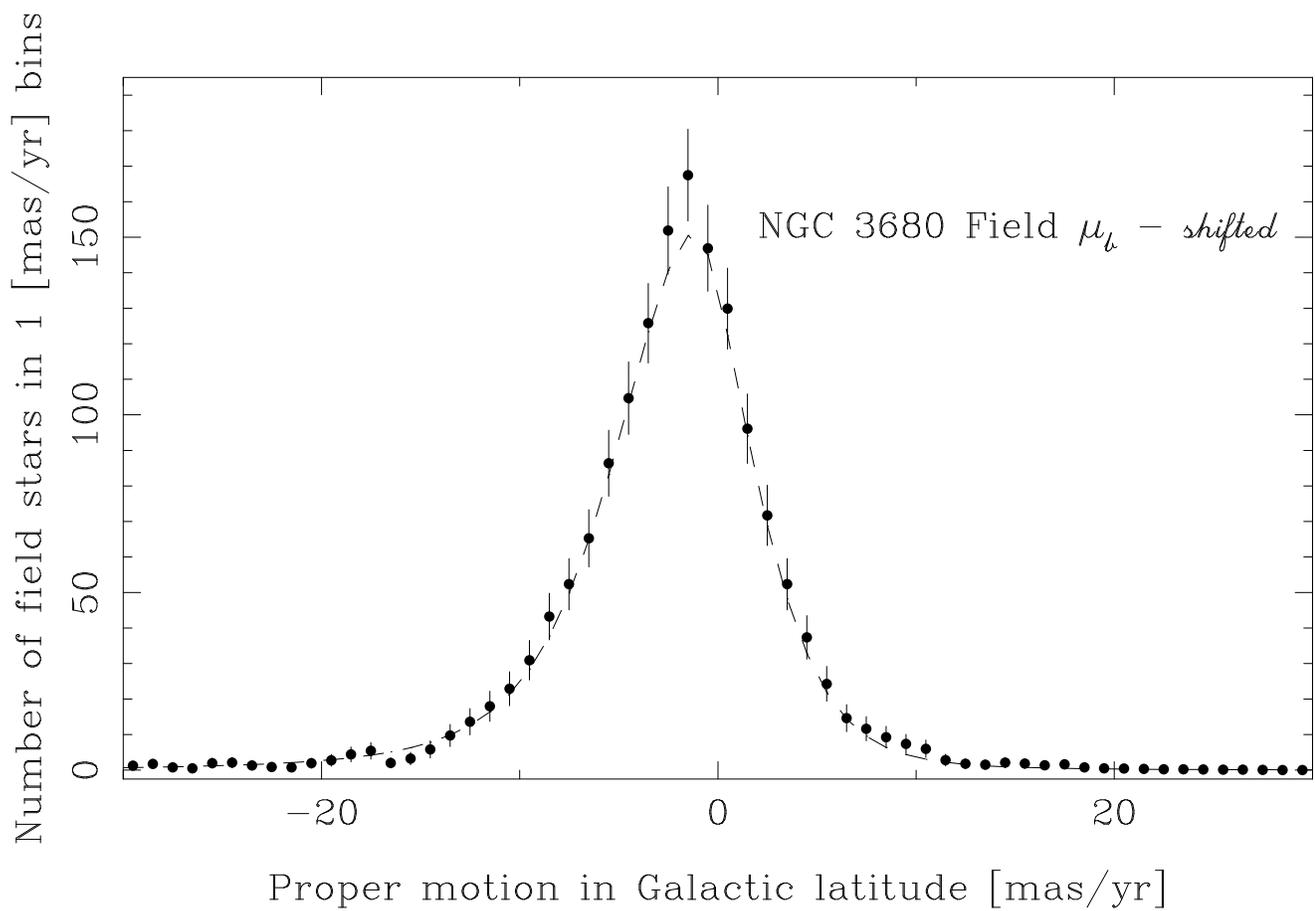
\end{document}